\documentclass[prl,twocolumn,amsmath,amssymb,groupedaddress,longbibliography]{revtex4-2}
\usepackage{bbm}
\usepackage{mathrsfs}
\usepackage{amsmath}
\usepackage{amsfonts}
\usepackage[colorlinks=true,citecolor=blue,anchorcolor=blue]{hyperref}
\usepackage{graphicx,epstopdf}
\usepackage{subfigure}
\usepackage{epsfig}
\usepackage{dcolumn}
\usepackage{bm}
\usepackage{color}
\usepackage{natbib}
\usepackage{amssymb}
\usepackage{xcolor}
\usepackage{braket}
\usepackage{soul}

\begin{document}

\title{Photonic scattering in 2D waveguide QED: Quantum Goos-Hänchen shift} 

\affiliation{Institute of Quantum Precision Measurement, State Key Laboratory of Radio Frequency Heterogeneous Integration, College of Physics and Optoelectronic Engineering, Shenzhen University, Shenzhen 518060, China}
\author{Yongguan Ke$^{1}$} \email{keyg@szu.edu.cn} 
\author{Zhenzhi Peng$^{1,2}$}
\author{Muhib Ullah$^{1}$}
\author{Chaohong Lee$^{1,3}$}\email{chleecn@szu.edu.cn}

\affiliation{$^{1}$Institute of Quantum Precision Measurement, State Key Laboratory of Radio Frequency Heterogeneous Integration, College of Physics and Optoelectronic Engineering, Shenzhen University, Shenzhen 518060, China}

\affiliation{$^{2}$Laboratory of Quantum Engineering and Quantum Metrology, School of Physics and Astronomy, Sun Yat-Sen University (Zhuhai Campus), Zhuhai 519082, China}

\affiliation{$^{3}$Quantum Science Center of Guangdong-Hong Kong-Macao Greater Bay Area (Guangdong), Shenzhen 518045, China}

\date{\today}

\begin{abstract}
Quantum emitters coupled to traveling photons in waveguides, known as waveguide quantum electrodynamics (WQED), offer a powerful platform for understanding light-matter interactions and underpinning emergent quantum technologies.
While WQED has been extensively studied in one dimension, two-dimensional (2D) WQED remains largely unexplored, where novel photonic scattering phenomena unique to higher dimensions are expected.
Here, we present a comprehensive scattering theory for 2D WQED based on the Green function method.
We show that the mean displacement between emitted and injected photons serves as a quantum analogue of the Goos-Hänchen shift.
When a photon is injected into a single off-centered port, the quantum Goos-Hänchen (QGH) shift can be enhanced in backward scattering under resonant conditions with subradiant states.
When a photon is injected into the center port, there is no QGH shift due to the mirror symmetry of structure.
However, for multiple-port injection with transverse momentum, the QGH shift is recovered and proportional to the derivative of phase with respect to transverse momentum. 
Unlike the classical Goos-Hänchen shift, these effects can be flexibly tuned by the injected photon's frequency.
Our work provides a general framework for exploring and manipulating photonic scattering in complex WQED networks.

\end{abstract}

\maketitle
\newpage

\emph{Introduction.}  Controlling the flow of light at the quantum level is essential for advancing quantum science and technology~\cite{pelucchi2022potential,Moody_2022}. 
In particular, platforms that enable coherent strong couplings between individual photons and localized quantum emitters are vital for quantum communication, quantum metrology, and scalable quantum computing~\cite{luo2023recent,Labonte2024}.
Among these platforms, waveguide quantum electrodynamics (WQED) has emerged as a powerful architecture, allowing engineered coupling between guided photons and spatially arranged emitters~\cite{RevModPhys.87.347,RevModPhys.89.021001,RevModPhys.90.031002,Sheremet2023}.
Implemented with superconducting qubits~\cite{van2013photon,brehm2021waveguide}, ultracold atoms~\cite{corzo2019waveguide,PhysRevLett.128.073601}, and quantum dots~\cite{tiranov2023collective}, WQED combines scalability and flexibility, offering a powerful interface between quantum optics and condensed matter physics.

Traditionally, WQED has focused on one-dimensional (1D) geometries, where many-body effects, photon-mediated interactions, and nonlinear quantum optics have been studied with high control~\cite{RevModPhys.87.347,RevModPhys.89.021001,RevModPhys.90.031002,Sheremet2023}.
However, generalizing WQED to two-dimensional (2D) architectures opens an entirely new regime of light–matter interaction~\cite{PhysRevLett.127.273602,PhysRevA.103.033702,PhysRevLett.132.163602}, where higher-dimensional scattering phenomena and geometric effects promisingly emerge.
In classical optics, Goos-H\"anchen shift describes the lateral mean position shift of reflected or transmitted light beams beyond the prediction of geometrical optics in two or three dimensions~\cite{goos1947neuer,lai2002large,kaiser1996resonances,Bliokh_2013,PhysRevApplied.12.014028}.
In recent years, the Goos-H\"anchen shift has been generalized to matter waves and potentially used for quantum enhanced sensing~\cite{Chen_2013,Le2025,Mckay2025}.
Despite these advances, a photonic analogue of the classical Goos-H\"anchen shift has remained elusive in the quantum regime, where individual photons interact with discrete quantum emitters.
%
%
Moreover, a general theoretical framework is urgently needed to explore geometric effects in 2D WQED.
However, in contrast to the extensively studied 1D WQED ~\cite{PhysRevLett.98.153003,PhysRevA.76.062709,PhysRevB.79.205111,PhysRevA.82.063816,PhysRevLett.108.143602,PhysRevLett.110.113601,PhysRevA.92.053834,Caneva_2015,manzoni2017simulating,Dinc2019exactmarkoviannon,PhysRevLett.122.073601,PhysRevLett.123.253601,PhysRevX.10.031011,PhysRevResearch.3.023030}, 2D architectures pose new challenges due to the increased complexity of emitter connectivity and the abundance of scattering channels.

\begin{figure}[!htp]
    \centering
    \includegraphics[width=0.51\textwidth]{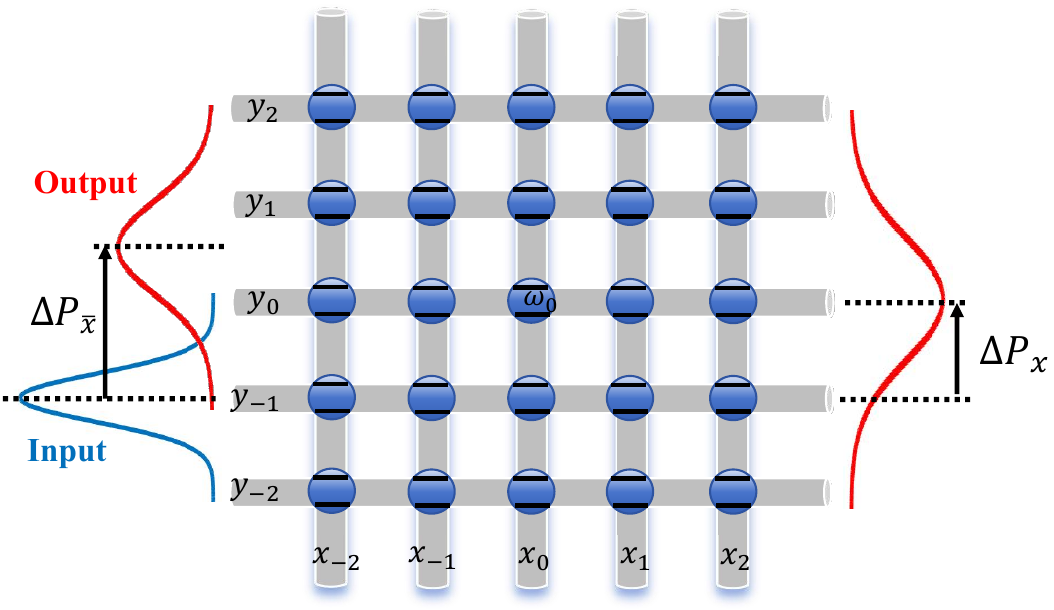}
    \caption{Schematics of a photon scattered by 2D atomic arrays coupled to waveguides. Two-level atoms with resonant frequency $\omega_0$ are placed in the nodes of crossing waveguides, and the waveguides are equally spaced. Quantum Goos-H\"anchen shift is manifested by the mean position shifts ($\Delta P_{x}$,$\Delta P_{\bar x}$)  in forward and backward scatterings, respectively.}
    \label{Schematics}
\end{figure}

In this letter, we develop a concise scattering theory for a single photon injected in 2D orthogonal waveguides coupled to atomic arrays at their nodes, as depicted in Fig.~\ref{Schematics}. 
The amplitudes in each port of forward, backward, upward and downward scatterings are related to the excitation Green function, which describes propagation of an excitation in the atomic network.
Considering a photon injected from $y_i$ port along the $x$ direction, the proportion between vertical and horizontal scatterings depends on the ratio of decay rates between the $y$ and $x$ directions.
We find the mean positions of photonic distribution in forward and backward scatterings can depart from $y_i$, which is dubbed as quantum Goos-H\"anchen (QGH) shift. 
The QGH shift depends on the injection ports and is enhanced by resonance with specific subradiant states in the backward scattering.
This is in stark contrast to the classical one, where no shift appears in backward scattering for vertical injection, let alone a port-dependent shift.
Under multiple-port excitation with transverse momentum, the QGH shift is proportional to the phase derivative with respect to transverse momentum in both forward and backward scattering and can be modulated by adjusting transverse momentum and injection frequency.
Our scattering theory readily applies to arbitrary 2D systems, pioneering a new direction in high-dimensional quantum photonics, where the interplay of geometry, topology, and strong light–matter coupling gives rise to unprecedented physical phenomena.

%

\emph{General formalism.}  Without loss of generality, we consider a photon propagating in 2D waveguide arrays, coupled to $N_x\times N_y$ quantum emitters at the nodes of crossings.
In the unit of $\hbar$, the system obeys the Hamiltonian,
\begin{equation}
\begin{split}
    H&=\sum_{s=\{x,y\},j}\int dk_s  c |k_s| a_{k_s,j}^\dag a_{k_s,j}+\sum_{j,l}  \omega_0 b_{j,l}^\dag b_{j,l} \\
    &+ \frac{ g_x}{\sqrt{2\pi}}\sum_{j,l} \int dk_x \big(a_{k_x,l}^\dag  e^{-ik_x x_j} b_{j,l}+h.c.\big)  \\
    &+ \frac{ g_y}{\sqrt{2\pi}}\sum_{j,l} \int dk_y \big(a_{k_y,j}^\dag  e^{-ik_y y_l} b_{j,l}+h.c.\big).
\end{split}
\end{equation}
Here, $a^\dag_{k_x,l} (a^{\dag}_{k_y,j})$ creates a photon with momentum $k_x$ ($k_y$) and frequency $c|k_x|$ ($c|k_y|$) propagating along the horizontal (vertical) waveguide at position $y_l$ ($x_j$),
and $b^\dag_{j,l}$ creates an excitation in the crossing of vertical and horizontal waveguides at the position ($x_j,y_l$).
The spacing between neighbor waveguides is set as $d$ in both the $x$ and $y$ directions, which is large enough to prohibit direct coupling between waveguides.
The relative frequency of excitation to the ground state is $\omega_0$, referred to as the resonant frequency.
When a photon in the horizontal (vertical) waveguide meets an atom, the propagating photon and the atomic excitation can be transferred to each other with a coupling strength $g_{x(y)}$.

Considering the scattering of a photon injected with longitudinal momentum $\kappa$ in superposition of multiple ports $y_l$, $|\psi_{i}\rangle=\sum_{l} f_{\kappa,l} a_{\kappa,y_{l}}^\dag|0\rangle$.
The scattering process can be simply understood as follows.
An injected photon is first transferred to an excitation, whose propagation in the atomic arrays is governed by the Green function
\begin{equation}
    G=(\omega-H_{eff})^{-1},
\end{equation}
with the effective Hamiltonian for the excitation
\begin{equation} \label{EffHam0}
\begin{split}
    H_{eff}&=\sum_{j,l}\omega_0 b_{j,l}^\dag b_{j,l}-i\Gamma_y \sum_{j,l,l'} b_{j,l}^{\dag}b_{j,l'}e^{i\omega/c|y_l-y_{l'}|} \\
    &-i\Gamma_x  \sum_{j,l,j'} b_{j,l}^{\dag}b_{j',l}e^{i\omega/c|x_j-x_{j'}|},
\end{split}
\end{equation}
where $\Gamma_{x(y)}=g_{x(y)}^2/c$ is the decay rate of an excitation along the $x(y)$ direction.
In addition to neighbor quantum emitters, an excitation may tunnel between remote quantum emitters and accumulates a phase dependent upon the distance.
Due to the open and lossy nature of the system~\eqref{EffHam0}, the radiative defects around the boundaries give rise to novel scale-free corner states, $|\varphi_{corner}(j,l)|^2 \sim (e^{-\alpha (j+l)/N}+e^{\alpha(j+l)/N})^2$.
Unlike conventional corner states, the decay length is inversely proportional to the system size $N_x=N_y=N$, which is scale-invariant~\cite{Supplementary}.

The excitation finally decays and emits a photon.
Its output amplitudes at different ports are determined by the Green function of excitation, the input amplitudes, and the system parameters~\cite{Supplementary},
\begin{equation} \label{EqScatter}
\begin{split}
    \chi_{x,l}&=f_{\kappa,{l}}-i\Gamma_x \sum_{j,j',l'} e^{-i\kappa {x_j}}e^{i\kappa x_{j'}} G_{j,l;j',l'} f_{\kappa,{l'}},  \\ 
   \chi_{\bar x ,l}&=-i\Gamma_x \sum_{j,j',l'} e^{i\kappa x_j}e^{i\kappa x_{j'}} G_{j,l;j',l'} f_{\kappa,{l'}},  \\ 
    \chi_{y,j}&=-i\frac{g_xg_y}{c} \sum_{j',l,l'} {e^{-i\kappa y_l}e^{i\kappa x_{j'}}} G_{j,l;j',l'} f_{\kappa,{l'}},  \\ 
    \chi_{\bar y,j}&=-i\frac{g_xg_y}{c} \sum_{j',l,l'} {e^{i\kappa y_l}e^{i\kappa x_j'}} G_{j,l;j',l'} f_{\kappa,{l'}}. 
\end{split}
\end{equation}
Here, $\chi_{x,j}$, $\chi_{\bar x,j}$ ($\chi_{y,j}$, $\chi_{\bar y,j}$) denote the amplitudes of a photon in forward, backward (upward, downward) scatterings in the $j$th waveguide along $x(y)$ direction, respectively.
We can alternatively calculate the scattering coefficients by using the transfer matrix method and obtain exactly the same results~\cite{Supplementary}. 
Compared to the transfer matrix method, the Green function method is superior in dealing with single-photon scattering in large and complex structures. 

\begin{figure}[!htp]
    \centering
    \includegraphics[width=0.5\textwidth]{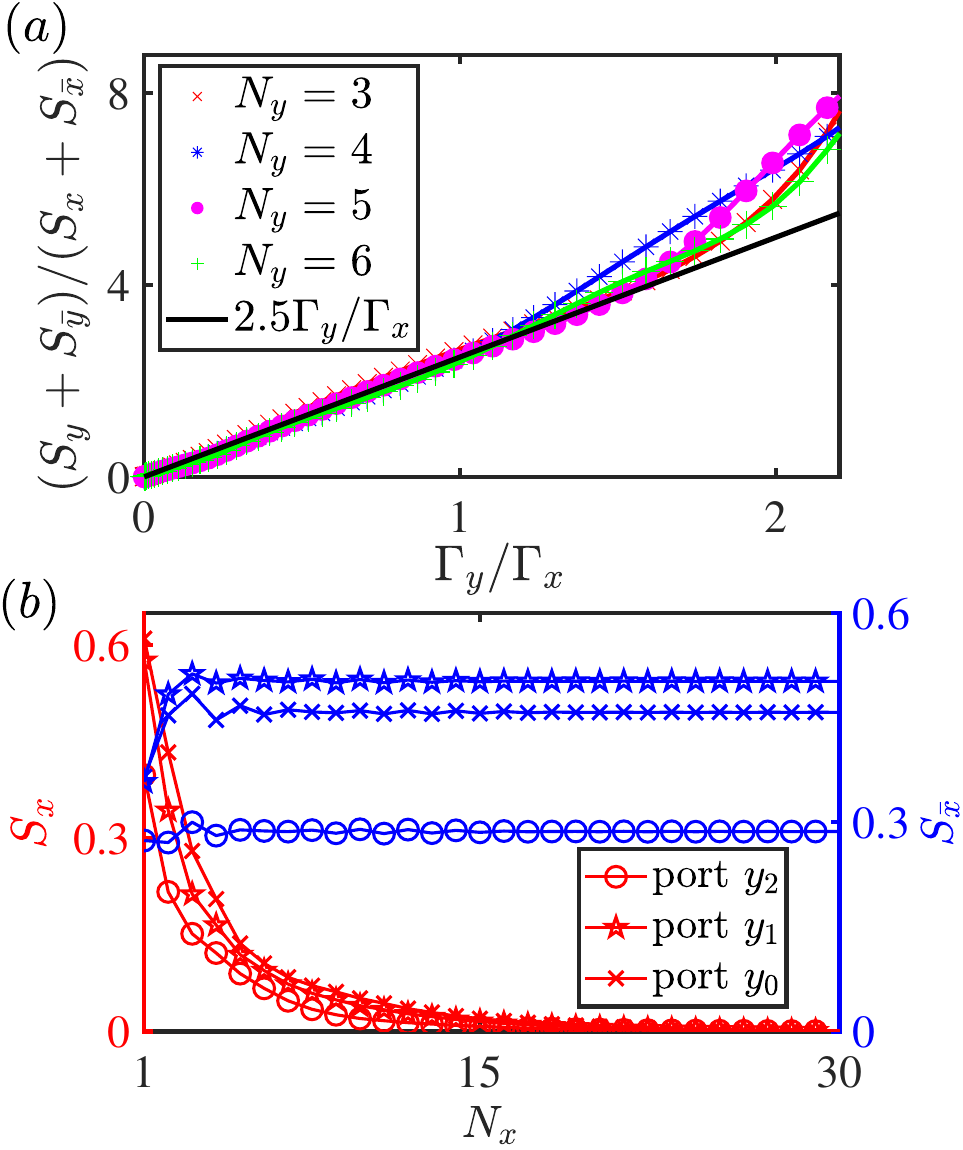}
    \caption{(a) The ratio of vertical and horizontal scatterings versus $\Gamma_y/\Gamma_x$. The frequency of injected photon is $\omega_0-0.078\Gamma_x$, the injection port is the first horizontal waveguide, and the other parameters are chosen as $d/c=\pi/(3\omega_0)$, $g_x=1$,  $c=100$, and $N_x=100$. Red `x’, blue `*', magenta solid dots, green `+' correspond to the numbers of waveguides along $y$ direction $N_y=3,4,5,6$, respectively. Black solid line denotes $(S_y+S_{\bar y})/(S_x+S_{\bar y}) \propto \Gamma_y/\Gamma_x$.  
    (b) The forward (red) and backward (blue) scatterings along $x$-direction versus $N_x$ for $N_y=5$ and different injection ports. `x’, star and cycle points denote the $y_0$, $y_1$ and $y_2$ ports, respectively. }
    \label{Distribution}
\end{figure}

\emph{Single-port injection.} 
We first consider the case of a single photon injected into a single port $y_{in}$ ($f_{\kappa, y_l}=\delta_{y_l,y_{in}}$). 
To understand the photonic scattering, we calculate the total probability from ports in both $x$ and $y$ directions,
\begin{eqnarray}
S_{v}&=&\sum_l |\chi_{v,l}|^2,~~v=\{x,y\}.
\end{eqnarray}
If the frequency of injected photon lies in the band gap of excitation, we find that the ratio between horizontal and vertical scattering probabilities $(S_y+S_{\bar y})/(S_x+S_{\bar x})$ is proportional to the ratio between the horizontal and vertical decay rates $\Gamma_y/\Gamma_x=(g_y/g_x)^2$; see Fig.~\ref{Distribution}(a).
For different numbers of layers along $y$ direction, $N_y$, these curves collapse to $\Gamma_y/\Gamma_x$ as $\Gamma_y/\Gamma_x \rightarrow 0$.
To understand the scaling law, we also calculate forward scattering $S_x$ and backward scattering $S_{\bar{x}}$ for different numbers of layers along the $x$ direction ($N_x$); see Fig.~\ref{Distribution}(b).
For different injection ports, all backward scatterings tend to be zero and the corresponding forward scatterings tend to be finite as $N_x$ increases.
Furthermore, the backward scattering decays with injected photonic frequency in the band gap and behaves damped oscillations in the continuous energy band~\cite{Supplementary}.  
Since the backward scattering is completely suppressed by the band gap~\cite{brehm2021waveguide}, 
according to the formula~\eqref{EqScatter}, we obtain a universal relation: $(S_y+S_{\bar y})/(S_x+S_{\bar x})=(S_y+S_{\bar y})/S_x\propto \Gamma_y/\Gamma_x$.

%
%
%
%
%

\begin{figure*}
    \centering
    \includegraphics[width=1\textwidth]{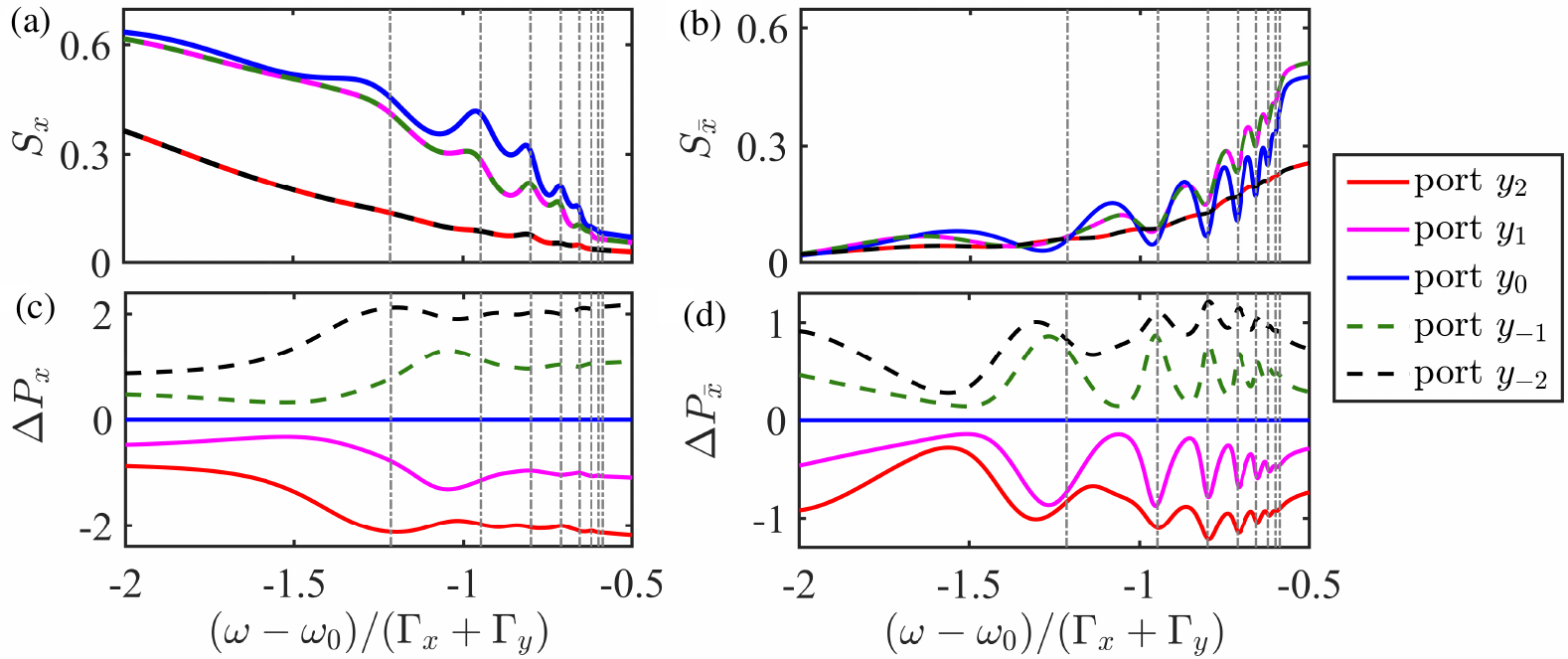}
    \caption{ Total probability and quantum Goos-H\"anchen shift in forward (a, c) and backward (b, d) scatterings for different injection frequencies and ports.
    The red solid, magenta solid, blue solid, green dashed, black dashed lines denote injection ports $y_2$, $y_1$, $y_0$, $y_{-1}$ and $y_{-2}$, respectively.
    The dashed-dot vertical dark lines indicate the frequencies of subradiant states.
    The parameters are chosen as $N_x=15$, $N_y=5$, $g_x=g_y=1$, $\omega_0 d/c=1$, and $\Gamma_x=\Gamma_y=0.01$.  
 }
    \label{Scattering}
\end{figure*}

To investigate how an underlying excitation affect forward and backward scattering of a photon, we analyze mean position displacements of forward and backward wavepackets relative to the injection port ($y_{in}$), 
\begin{equation}
    \Delta P_{x(\bar x)}=P_{x(\bar x)}-P_{in},
\end{equation}
where the mean positions along $s$-direction  are given by $P_{s}={\sum_l l|\chi_{s,l}|^2}/{\sum_l |\chi_{s,l}|^2}$ with $s=\{x,\bar x\}$ and $P_{in}=\sum_l |f_{\kappa,l}|^2 l$, respectively.
Figures~\ref{Scattering}(a)-(d) display total probabilities and mean position shifts of forward and backward scatterings for different injection frequencies and ports.    
Our simulations are performed for a square atomic array with isotropic coupling strengths $g_x=g_y=1$ and decay rates $\Gamma_x=\Gamma_y=0.01$.
Because the system has mirror symmetry in $y$-direction, the total probabilities of forward and backward scatterings from the injection port $y_s$ are the same as those from the injection port $y_{-s}$~\cite{Supplementary}. 

Away from resonance, the photon barely interacts with the atomic array and propagates transparently, leading to dominant forward transmission and negligible backward reflection.
In contrast, rich phenomena occur around the resonant frequencies $\omega_0$, where the excitation spectrum of the atomic array supports subradiant modes~\cite{Supplementary}.
The forward (backward) scattering is enhanced (diminished) around some frequencies marked by dashed vertical lines, which correspond to the subradiant states with eigenvalues $\omega_x^{s}+\omega_y^{s_0}$ in Markov approximation ($\omega c/d\approx \omega_0 c/d$)~\cite{PhysRevLett.123.253601}. 
Here, $\omega_x^{s}$ is the eigenvalue of subradiant states in a one-dimensional WQED along $x$-direction, while $\omega_y^{s_0}$ is the eigenvalue of the most subradiant state in a one-dimensional WQED along $y$-direction. 
We find that the mean position is shifted from the injection port in both forward and backward scatterings, dubbed as quantum Goos-H\"anchen (QGH) shift.
The magnitudes and signs of the QGH shifts strongly depend on the injection port.
Due to mirror symmetry, a photon injected from the center port is distributed symmetrically, leading to no QGH shift.
For off-center injections, the QGH shifts corresponding to injection ports $y_s$ and $y_{-s}$ are equal in magnitude but opposite in sign [Figs.~\ref{Scattering}(c, d)], as required by mirror symmetry~\cite{Supplementary}.
In particular, the QGH shift in backward scattering increases when the injected photon is in resonance with the subradiant states. 
This is in stark contrast to the classical GH shift~\cite{goos1947neuer,lai2002large,kaiser1996resonances,PhysRevApplied.12.014028}, in which (i) the center of reflected light beam has no shift when the injected light beam is perpendicular to the medium interface and (ii) the transmission and reflection behaviors are independent of the injected position.
Furthermore, the QGH shift is comparable to the waveguide spacing, which can be designed as large as possible.

\begin{figure}[!htp]
    \centering
    \includegraphics[width=0.5\textwidth]{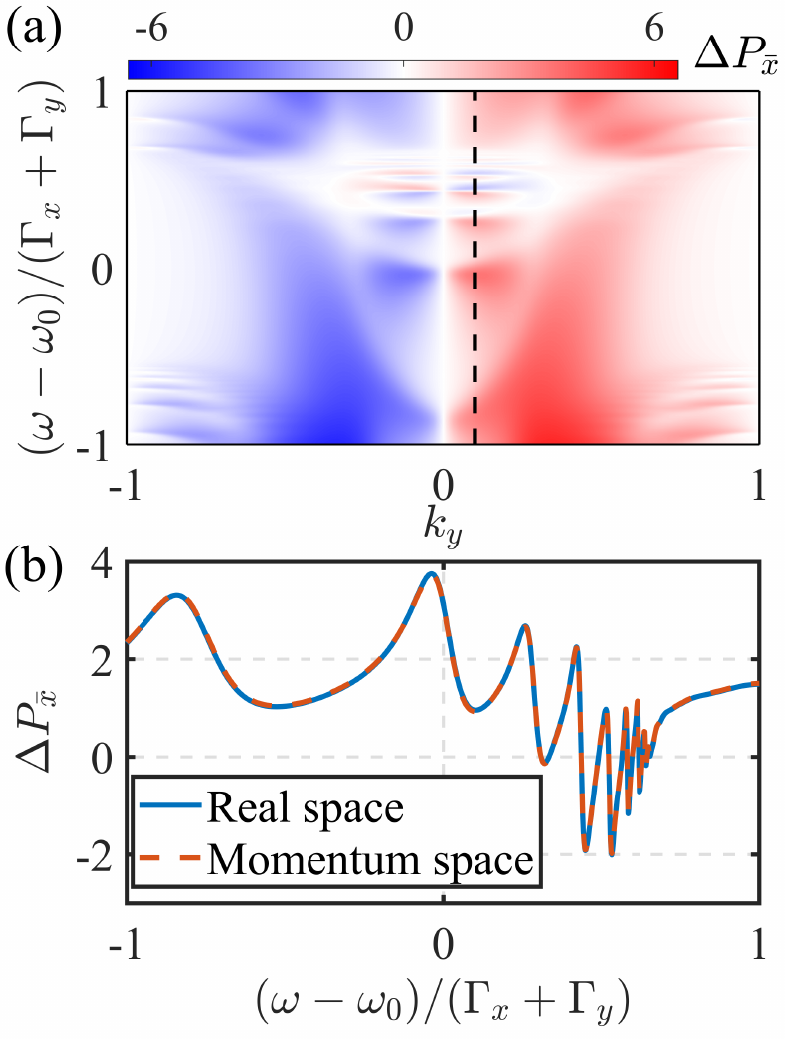}
    \caption{(a) The quantum Goos–Hänchen shift of reflected photon as a function of injection frequency and transverse momentum of Gaussian wavepacket. (b) The quantum Goos–Hänchen shift as a function of injection frequency with transverse momentum $k_y=0.1\pi$. All calculations are performed with $g_x=g_y=1$,  $\omega_0d/c=1$, $N_x=15$, and $N_y=25$. }
    \label{QGHshift_Gaussian}
\end{figure}

\emph{Multiple-port injection.} As discussed above, due to mirror symmetry, there is no QGH shift when the photon is injected into the middle waveguide. 
However, by using multiple-port injection, we can recover QGH shift even when the wavepacket profile of the injected photon centers in the middle waveguide. 
Assume the initial input $|\psi_{i}\rangle=\sum_{l} f_{\kappa,l} a_{\kappa,y_{l}}^\dag|0\rangle$ as a Guassian wavepacket, that is, 
\begin{equation}
f_{\kappa, l}=\frac{e^{-\frac{l^2}{4\sigma^2}+ik_y l}}{(2\pi)^{1/4}\sqrt{\sigma}},
\end{equation}
where $\sigma$ is the width of the wavepacket and $k_y$ is the transverse wave vector.
Without loss of generality, we calculate the QGH shift in backward scattering for different injection frequencies and transverse wave vectors; see Fig.~\ref{QGHshift_Gaussian}(a).
Similar to the case of single-port injection, there is no QGH shift when the transverse wave vector is $0$ due to mirror symmetry.
However, non-zero transverse wave vectors, corresponding to non-zero transverse momenta $\hbar k_y$, play a crucial role in realizing the QGH shift.
The QGH shift becomes opposite when the transverse momentum changes from $+k_y$ to $-k_y$.
It is natural to expect that the positive (negative) transverse momentum leads to positive (negative) QGH shift.
However, it is still possible that the QGH shift can be opposite to the transverse momentum for some injection frequencies; see Fig.~\ref{QGHshift_Gaussian}(b).

To understand the negative QGH shift, we transfer to the momentum space to derive a formula of QGH shift.
First, we respectively perform a Fourier transform on both the injected and reflected fields, 
\begin{eqnarray}
h_{in}(k_y)&=&\sum_{l}f_{\kappa,l}e^{-i k_y l}, \nonumber \\
h_{\bar x, re}(k_y)&=&\sum_{l}\chi_{\bar x, l}e^{-i k_yl}. \nonumber 
\end{eqnarray}
Then, to connect the reflected field to the injected one, we rewrite the reflected field as 
\begin{equation}
h_{\bar x, re}(k_y)=h_{in}(k_y) r_{k_y} e^{i\theta_{k_y}},
\end{equation}
where $r_{k_y}$ and $\theta_{k_y}$ are the relative amplitude and phase.
At last, the QGH shift can be given as, 
\begin{equation} \label{QGHMomentumSpace}
\Delta P_{\bar x}=-\frac{\sum_{k_y} |h_{\bar x, re}(k_y)|^2 {\partial_{k_y} \theta_{k_y}}}{\sum_{k_y} |h_{\bar x, re}(k_y)|^2},
\end{equation}
which depends on the derivative of relative phase with respect to the transverse wave vector.
The QGH shift in the momentum-space representation yields results identical to those in the real-space representation, as shown by the red dashed and blue solid lines in Fig.~\ref{QGHshift_Gaussian}(b), respectively. 
From Eq.~\eqref{QGHMomentumSpace}, we deduce that the relative phase increasing with the transverse wave vector around the center wave vector leads to a negative QGH shift.
In addition to adjusting the injection frequency and port, the QGH shift can also be flexibly tuned by controlling the transverse momentum.
This indicates that the 2D WQED offers a versatile platform to control QGH shift.

\emph{Discussions.} Using Green functions, we have developed a general theoretical framework for single-photon scatterings in two-dimensional atomic arrays coupled to orthogonal waveguides. 
This method offers significant advantages for treating large-scale or complex structures, where traditional transfer-matrix approaches become analytically and numerically intractable. 
By relating scattering amplitudes to the excitation dynamics within the emitter array, our approach provides both physical transparency and computational versatility. 
Using this formalism, we have uncovered a quantum Goos–H\"anchen shift in backward scattering resonantly enhanced by subradiant states. 
These effects reveal a rich landscape of directional quantum scattering in 2D waveguide quantum electrodynamics, where light transport is fundamentally shaped by quantum interference and geometry.

Beyond square lattices, our Green function framework can be extended to nontrivial geometries and engineered gauge fields, facilitating the exploration of topological phases and their impact on quantum light propagation~\cite{hallacy2025nonlinear,Perczel2020,PhysRevLett.131.103604,hallacy2025nonlinear}. 
It further enables studies of Anderson (many-body) localization in photonic scattering for disordered atomic arrays beyond 1D~\cite{PhysRevResearch.3.033233}. 
The framework is also generalizable to the few-photon regime: akin to earlier studies of inelastic scattering in 1D waveguide quantum electrodynamics~\cite{PhysRevA.82.063816,PhysRevLett.123.253601}, our method can be a benchmark to address few-photon dynamics in 2D networks, with applications in quantum simulation, transport, and non-equilibrium photonics. 
Altogether, this work establishes a versatile theoretical foundation for 2D waveguide quantum electrodynamics, bridging quantum optics, condensed matter physics, and photonic quantum technologies.

\begin{acknowledgments}
We acknowledge useful discussions with Yuri S. Kivshar, Feng Wu, Wenjie Liu, Ling Lin and Li Zhang.
This work is supported by the National Natural Science Foundation of China (Grants No. 12025509, 12275365, and 92476201), the National Key Research and Development Program of China (Grant No. 2022YFA1404104), the Guangdong Provincial Quantum Science Strategic Initiative (GDZX2305006 and GDZX2405002), and the Natural Science Foundation of Guangdong Province (Grant No. 2023A1515012099).
\end{acknowledgments}


\providecommand{\noopsort}[1]{}\providecommand{\singleletter}[1]{#1}%

\onecolumngrid
\clearpage

\renewcommand {\Im}{\mathop\mathrm{Im}\nolimits}
\renewcommand {\Re}{\mathop\mathrm{Re}\nolimits}
\renewcommand {\i}{{\rm i}}
\renewcommand {\phi}{{\varphi}}

\begin{center}
	\noindent\textbf{\large{Supplemental Materials of}}
	\\\bigskip
	\noindent\textbf{\large{``Photonic scattering in 2D waveguide QED: Quantum Goos-Hänchen shift"}}
	\\\bigskip
	\onecolumngrid
	
	Yongguan Ke$^{1,*}$, Zhenzhi Peng$^{1,2}$, Muhib Ullah$^{1}$, Chaohong Lee$^{1,3 \dag}$
	
	\small{$^1$ \emph{Institute of Quantum Precision Measurement, State Key Laboratory of Radio Frequency Heterogeneous Integration, College of Physics and Optoelectronic Engineering, Shenzhen University, Shenzhen 518060, China}}\\
	\small{$^2$ \emph{Laboratory of Quantum Engineering and Quantum Metrology, School of Physics and Astronomy, Sun Yat-Sen University (Zhuhai Campus), Zhuhai 519082, China}}\\
	\small{$^3$ \emph{Quantum Science Center of Guangdong-Hong Kong-Macao Greater Bay Area (Guangdong), Shenzhen 518045, China}}\\
\end{center}

\setcounter{equation}{0}

\setcounter{figure}{0}

\setcounter{table}{0}
\renewcommand{\theequation}{S\arabic{equation}}

\renewcommand{\thefigure}{S\arabic{figure}}
\renewcommand{\theHfigure}{S\arabic{figure}}

\renewcommand{\thesection}{S\arabic{section}}


\section{S1. Scale-free localized corner states}\label{Localized_States}

%
%
In  a 1D WQED with a unit cell containing two atoms, we have also found scale-free localized edge states, whose energies are distributed in two inverse energy bands for trivial phase and one inverse energy band for topological phase~\cite{PhysRevLett.131.103604New}.
Interestingly, we find scale-free localized corner states in the 2D WQED of a square lattice, which can be a natural generalization of the scale-free localized edge states. 
In this section, we will analytically explain the origin of scale-free localized corner states, and figure out the  distribution of scale-free localized corner states in inverse energy.

Observing that the couplings along one direction do not depend on the other direction, we can decouple the effective Hamiltonian as 
\begin{equation}
    H_{eff}=H_x^{1D}\otimes I_y+I_x\otimes H_y^{1D},
\end{equation}
where $H_{x(y)}^{1D}$ and $I_{x,y}$ are the effective Hamiltonian  and the identity matrix in the $x(y)$ direction. 
Adopted the similar analysis from Ref.~\cite{Yao2021New}, once we obtain the eigenstates $|\psi_m^{x(y)}\rangle$ and eigenvalues $\omega_m^{x(y)}$ of the 1D WQED, we can immediately obtain the eigenstates and eigenvalues of the 2D WQED, that is, $|\psi_{m,n}\rangle=|\psi_m^{x}\rangle\otimes |\psi_n^{y}\rangle$ with $\omega_{m,n}=\omega_{x}^{m}+\omega_{y}^{n}$.
From Ref.~\cite{PhysRevLett.131.103604New}, we have already known that $H_{x(y)}^{1D}$ and $(H_{x(y)}^{1D}-\omega_0)^{-1}$ share the same eigenstates, it is natural to define an inverse Hamiltonian for such 2D square lattice as 
\begin{equation} \label{HamInv2DSquare}
    H_{inv}^{2D}=(H_x^{1D}-\omega_0)^{-1}\otimes I_y+ I_x\otimes (H_y^{1D}-\omega_0)^{-1},
\end{equation}
which shares the same eigenstates with the effective Hamiltonian (3) in the main text.
Because the eigenvalues of $(H_{x(y)}^{1D}-\omega_0)^{-1}$ are continuous functions of $k_x(k_y)$, we can deduce that the inverse Hamiltonian~\eqref{HamInv2DSquare} can also give continuous inverse energy bands.

Back to scale-free localized corner states in a finite square lattice, analyzing the inverse Hamiltonian~\eqref{HamInv2DSquare} is more convenient for some analytical results.
The key is to find the eigenstates and eigenvalues of $(H_{x(y)}-\omega_0)^{-1}$.
%
%
Without loss of generality, we consider that the 2D QED system is homogenous and $H_x^{1D}$ and $H_y^{1D}$ have the same radiative decay.
We first numerically find that $(H_{x}^{1D}-\omega_0)^{-1}$ corresponds to a discrete tight-binding model, which contains nearest-neighboring couplings and onsite potential. 
This model can be written in the following form~\cite{PhysRevLett.131.103604New,RevModPhys.95.015002New},
\begin{align}\label{31}
    (H_x^{1D}-\omega_0)^{-1}&=\sum_{j=1}^{N-1}[C(b_{j+1}^{\dag} b_j +b_j^{\dag} b_{j+1})]+\sum_{j=2}^{N-1}Bb_j^{\dag} b_j+A(b_1^{\dag} b_1+b_N^{\dag} b_N),
\end{align}
where $C$ denotes nearest-neighboring hopping strength, $B$ and $A$ denote onsite potential in the bulk and at the boundaries, respectively. 
According to the inverse relation, $(H_{x}^{1D}-\omega_0)^{-1}(H_{x}^{1D}-\omega_0) = I_x$, we find that $A$, $B$, $C$ satisfy,
\begin{equation}\label{ABC}
\begin{aligned}
   & -i\Gamma_x(A+Ce^{i\varphi})=1,\\
    &-i\Gamma_x(Ae^{i\varphi}+C)=0, \\
    &-i\Gamma_x(2Ce^{i\varphi}+B)=1.       
\end{aligned}
\end{equation}
By solving Eq.~\eqref{ABC}, we can analytically obtain the relation between the parameters of $(H_x^{1D}-\omega_0)^{-1}$ and  $H_x^{1D}$, that is, $A=-\frac{1}{2\Gamma_x}({\rm cot}\varphi-i)$, $B=-\frac{{\rm cot}\varphi}{\Gamma_x}$, and  $C=\frac{1}{2\Gamma_x{\rm sin}\varphi}$.       
The complex onsite potential $A$ corresponds to radiative decay at the boundaries.
The non-Hermitian Hamiltonian~\eqref{31} can support scale-free localized edge states. 
Once the scale-free localized edge states are obtained, we can immediately construct scale-free localized corner states. In the following, we will show how to analytically derive the scale-free localized edge states.  

\begin{figure}
    \centering
\includegraphics[width=0.98\textwidth]{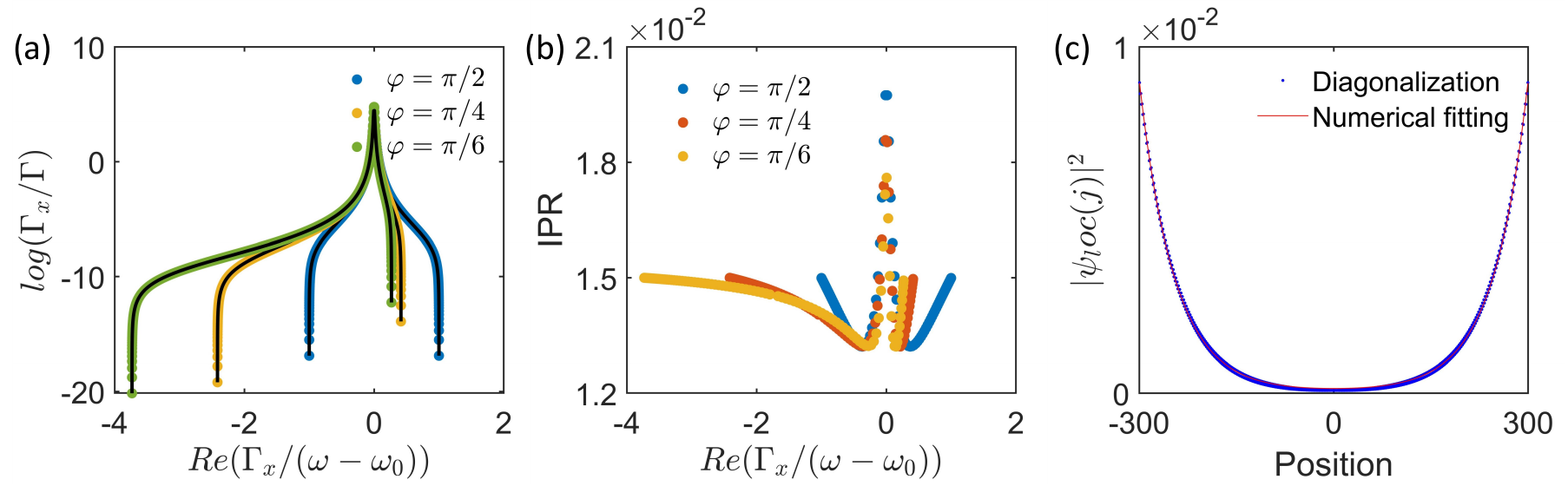}
    \caption{(a) The relation between logarithm of the decay rate and inverse energy. Here, the dots represents eigenvalues obtained from exact diagonalization, and the black lines are given by Eq.~\eqref{s28}. Blue, yellow, green dots correspond to $\varphi=\pi/2$, $\pi/4$, $\pi/6$, respectively.  (b) inverse participation ratio of eigenstates in different value of $\varphi$. (c) Symmetric scale-free localized states with the largest IPR (blue dots) with $\varphi=\pi/2$, and the fitting function (blue line) $|\psi_{loc}(j)|^2 \sim (e^{-\alpha F/N}+e^{\alpha F/N})^2$. The other parameters are chosen as $g_x=1$, $c=100$, and $N=600$. }\label{af1}
\end{figure}

We assume that the scale-free localized edge states $|\psi\rangle= \sum_j\psi_jb_j\dag |0\rangle$ satisfy~\cite{PhysRevB.108.L161409New}
\begin{equation}\label{36}
    \psi_j(\beta)=R\beta^j+T\beta^{-j},
\end{equation}
where $\beta$ is a complex number, and $R$, $T$ are coefficients determined by the boundary condition.
We can obtain eigenvalue equations 
\begin{eqnarray}\label{35}
    C(\psi_{j-1}+\psi_{j+1})+B\psi_j=E\psi_j
\end{eqnarray}
in the bulk, and
\begin{equation}\label{38}
    \begin{aligned}
        C\psi_2+A\psi_1=&E\psi_1,\\
        C\psi_{N-1}+A\psi_N=&E\psi_N
        \end{aligned}
\end{equation}
at the boundaries.
Substituting the ansatz~\eqref{36} into Eq.~\eqref{35}, we can find
\begin{eqnarray}\label{mid_change}
    C(\psi_{j-1}+\psi_{j+1})&=&C(R\beta^{j-1}+T\beta^{1-j}+R\beta^{j+1}+T\beta^{-j-1})\nonumber\\
    &=&\beta[C(R\beta^j+T\beta^{-j})]+\beta^{-1}[C(R\beta^{j}+T\beta^{-j})]\nonumber\\
    &=&C(\beta+\beta^{-1})\psi_j.
\end{eqnarray}
Combining Eqs.\eqref{35} and \eqref{mid_change}, we can obtain the bulk equation
\begin{eqnarray}\label{37}
    C(\beta+\beta^{-1})+B=E.
\end{eqnarray}
Substituting the ansatzs~\eqref{36} and \eqref{37} into Eqs.~\eqref{38}, we can find
\begin{align}\label{39}
        &\left[(A-B)\beta-C\right]R+\left[(A-B)\beta^{-1}-C\right]T=0,\\
         &\left[(A-B)\beta^N-C\beta^{N+1}\right]R+\left[(A-B)\beta^{-N}-C\beta^{-(N+1)}\right]T=0.\nonumber
    \end{align}
The nonzero solution requires that the determinant of the coefficients is equal to zero, which means $\beta$ satisfies
\begin{equation}\label{40}
    \left[(A-B)\beta-C\right]\left[(A-B)\beta^{-N}-C\beta^{-(N+1)}\right]=\left[(A-B)\beta^{-1}-C\right]\left[(A-B)\beta^N-C\beta^{N+1}\right],
\end{equation}
The scale-free localization of eigenstates implies that $\beta$ takes the form,
\begin{equation}\label{41}
    \beta=e^{F/N+i\theta},
\end{equation}
where $\theta\in \left[0,2\pi \right]$ and $F$ is a real number. Combining Eq.~\eqref{40} and Eq.~\eqref{41}, we can obtain the relation between $F$ and original Hamiltonian coefficients
\begin{equation}\label{42}
    e^{2F}=(\frac{(A-B)e^{i\theta}-C}{(A-B)e^{-i\theta}-C})^2e^{-2i(N+1)}.
\end{equation}
By taking absolute value on both side, the value of $F$ is given by
\begin{equation}\label{43}
    F=\frac{1}{2}ln|(\frac{Je^{i\theta}-C}{Je^{-i \theta}-C})^2|.
\end{equation} 
By substituting Eq.~\eqref{41} into $C(\beta+\beta^{-1})+B=E$, the
eigenvalues can be decomposed into real and imaginary parts
\begin{equation}\label{44}
    (\text{Re}(E),\text{Im}(E))=(2C{\rm cos}\theta+B,2C{\rm sin}\theta F/N).
\end{equation}
Then, we can obtain the relation between inverse energy and decay rate,
\begin{equation}\label{s28}
    (\frac{\Gamma_x}{\omega-\omega_0},\frac{\Gamma}{\Gamma_x})=(2C{\rm cos}\theta+B,\frac{2C{\rm sin}\theta F/N}{(2C{\rm cos}\theta+B)^2+(2C{\rm sin}\theta F/N)^2}).
\end{equation}
We compare the eigenvalues obtained by Eq.~\eqref{s28} and the numerical exact diagonalization through fixing $N=600$ and varying different phases $\varphi=\pi/6,~\pi/4,~\pi/2$ in Fig.~\ref{af1}(a).
The other parameters are chosen as $g_x=1$, $\omega_0=100$, and and $c=100$.
The black lines denote the analytical results and the dots denote the numerical results. 
We can find that these analytical results are consistent with the numerical results.
The most superradiant states are concentrated at  inverse energy of zero with different phase constant $\varphi$. 
To search for scale-free localized states effectively, we calculate inverse participation ratio of eigenstates in a finite array under open boundary condition
\begin{eqnarray}\label{45}
    \text{IPR}=\frac{\sum_j|\psi_j|^4}{(\sum_j|\psi_j|^2)^2}.
\end{eqnarray}
 The IPR tends to 1 for the most localized state and 0 for the extended state, which can be used to distinguish localized and extended states. 
 Fig.~\ref{af1}(b) shows IPR as a function of the inverse energy for different phases $\varphi=\pi/6,~\pi/4,~\pi/2$. 
 Interestingly, states with larger $\text{IPR}$ are also distributed around the zero inverse energy, indicating that these localized states are superradiant states. 
 To show the scale-free properties of the localized states, we show spatial distribution of the localized state with the largest $\text{IPR}$ in Fig.~\ref{af1}(c).
 The dots are obtained by numerical exact diagonalization and the solid line is fitted by $|\psi_{loc}(j)|^2 \sim (e^{-\alpha F/N}+e^{\alpha F/N})^2$, which are consistent with each other.
 We confirm that the localized states are scale-free localized edge states.
 Because an excitation localized around the boundaries has larger chance to escape from the quantum emitters, the scale-free localized edge states have larger decay rate and turn out to be superradiant states.

\begin{figure*}
    \centering
    \includegraphics[width=0.99\textwidth]{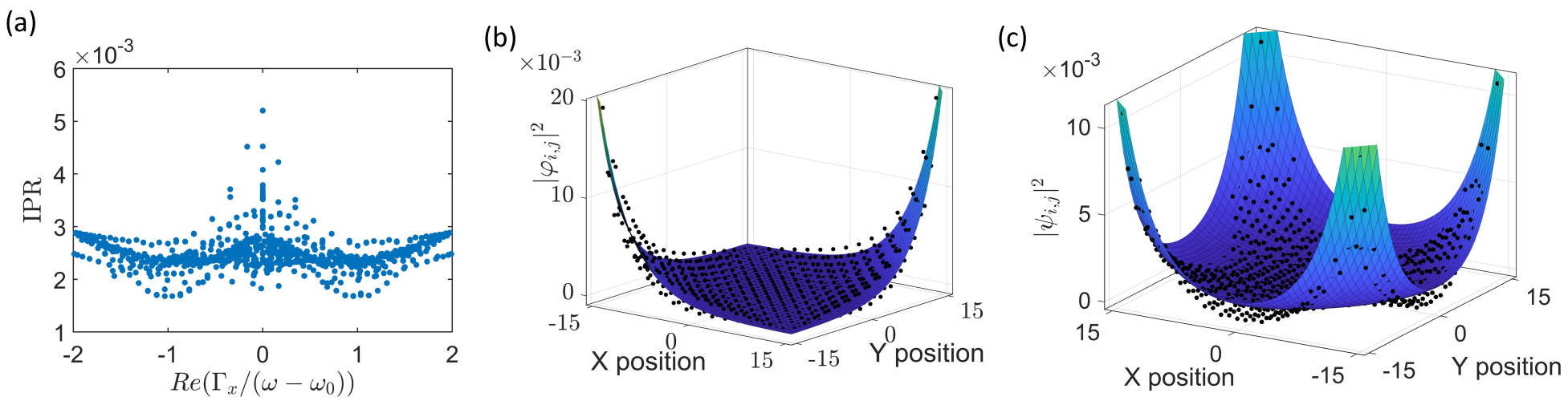}
    \caption{(a) Inverse participation ratio of eigenstates as a function of real part of inverse energy. (b) Symmetric scale-free localized state with $E^{-1}\approx-0.085+0.1835i$  and the largest IPR(black dot), and the fitting function $|\varphi_{loc}(i,j)|^2 \sim (e^{-F(i+j)/N}+e^{F(i+j)/N})^2$. (c) The spatial distribution of quantum state $|\varphi_{i,j}|^2$ with $E^{-1}\approx-0.085+0.1835i$ and the $\text{No}.~26$ largest IPR,  and the fitting function $|\varphi_{loc}(i,j)|^2 \sim (e^{-F(i+j)/N}+e^{F(i+j)/N}+e^{-F(i-j)/N}+e^{F(i-j)/N})^2$.
    The above calculations are performed with $N_x=N_y=30$, $c=100$, $g_x=1$, $\Gamma_x=0.01$, $\lambda=1$, and $\varphi=\pi/2$}\label{f3}
\end{figure*}

Since we have already analytically obtained the 1D inverse Hamiltonian and the scale-free localized edge states, It is natural to derive the 2D inverse Hamiltonian, 
\begin{eqnarray}\label{46}
{H_{inv}^{2D}}&=&\sum_{i,j}C(b_{i,j+1}^{\dag} b_{i,j} +b_{i+1,j}^{\dag}b_{i,j}+h.c.)+\sum_{ bulk }B b_{i,j}^{\dag} b_{i,j}
+\sum_{edge} A b_{i,j}^{\dag} b_{i,j}
+\sum_{corner} 2A b_{i,j}^{\dag}  b_{i,j}. \nonumber 
\end{eqnarray}
Because the 2D inverse Hamiltonian inherits the radiative loss from the 1D inverse Hamiltonian, this Hamiltonian is also non-Hermitian with radiative decay at four edges and four corners. 
The decay rate at the corner is twice that at the edge, because there are one and two escape ports connecting to the atoms at the edge and corner, respectively.
The radiative decays at the edges and corners give rises to scale-free localized corner states.
As products of scale-free localized edge states in the $x$ and $y$ directions, the scale-free localized corner states have inverse energies around $0$.
To diagnose scale-free localized corner states,
we calculate the ratio of inverse participation as a function of inverse energies; see Fig.~\ref{f3}(a).
We indeed find some corner states with large IPR around $0$ inverse energy; see Figs.~\ref{f3}(b, c).
There are two localization patterns, one is the localization at two corners, and the other is localization at all four corners.
Fig.~\ref{f3}(b) shows the probability distribution of the most localized state, satisfying the first localization pattern.
We try to fit the most localized state with the form of scale-free localized function, $|\varphi_{loc}(i,j)|^2 \sim (e^{-F(i+j)/N}+e^{F(i+j)/N})$; see black dots in Fig.~\ref{f3}(b).
The fitting function agrees well with the exact state.
The first localization pattern comes from the superposition of symmetric and asymmetric product states, which leads to destructive interference of the other two corners. 
Fig.~\ref{f3}(c) presents the spatial distribution of the quantum state with the $\text{No}.~26$ largest IPR, satisfying the second localized pattern which can be well described by the function $|\varphi_{loc}(i,j)|^2 \sim (e^{-F(i+j)/N}+e^{F(i+j)/N}+e^{-F(i-j)/N}+e^{F(i-j)/N})^2$.
%

%

\section{S2. Green function method}
Here, we give the details of derivation of photonic scattering of 2D WQED.
We first consider the photonic scattering in 2D WQED under open boundary condition. 
We can split the Hamiltonian into two parts, $H=H_0+V$, with $H_0$ being the energy of a bare photon and excitation, and $V$ being the interaction between excitation and photon.
According to the Lippmann-Schwinger equation, given an initial state $|\psi_{i}\rangle=\sum_{l} f_{\kappa,l} a_{\kappa,y_{l}}^\dag|0\rangle$, the output state reads
\begin{equation}
    |\psi_{o}\rangle= |\psi_{i}\rangle+\frac{1}{E-H_0}V|\psi_{o}\rangle.
\end{equation}
Assume that the output state takes the form of
\begin{equation}
|\psi_{o}\rangle =\sum_{j,l} Q_{j,l}b_{j,l}^\dag |0\rangle+\sum_{l}\int dk_x O_{k_x,l}a_{k_x,l}^\dag|0\rangle+\sum_{j}\int dk_y W_{k_y,j}a_{k_y,j}^\dag|0\rangle,
\end{equation}
where $Q_{j,l}$, $O_{k_x,l}$, and $W_{k_y,j}$ are the amplitudes of excitation at the $(j,l)$th quantum emitters,  photon propagating along the $l$th waveguide in the horizontal direction and the $j$th waveguide in the vertical direction, respectively.
Substituting the above ansatz into the Lippmann-Schwinger equation, we can obtain relations between $Q_{i,j}$, $O_{k_x,l}$ and $W_{k_y,j}$,
\begin{eqnarray} \label{Coupling} 
  &&  Q_{j,l}=\frac{g_x}{\sqrt{2\pi}}\int dk_x \frac{e^{ik_x x_j }O_{k_x,l}}{\omega-\omega_{0}} +\frac{g_y}{\sqrt{2\pi}} \int dk_y \frac{e^{ik_y y_l }W_{k_y,j}}{\omega-\omega_{0}},  \nonumber \\
   && O_{k_x,l}= \delta_{k_x,\kappa}f_{\kappa,l}+ \frac{g_x}{\sqrt{2\pi}} \sum_{j} \frac{e^{-ik_x x_j}}{\omega-\omega_{k_x}} Q_{j,l},  \\
  && W_{k_y,j}= \frac{g_y}{\sqrt{2\pi}} \sum_{l} \frac{e^{-ik_y y_l}}{\omega-\omega_{k_y}} Q_{j,l}.\nonumber
\end{eqnarray}
By eliminating the freedom of photon ($O_{k_x,j}$, $W_{k_y,l}$), we can obtain the motion equation for excitation
\begin{equation}
    Q_{j,l}=\frac{g_x}{\sqrt{2\pi}}\frac{e^{i\kappa x_j}}{\omega-\omega_0}f_{\kappa,l}-i\Gamma_x \sum_{j'}\frac{e^{i \omega/c |x_j-x_j'|}}{\omega-\omega_0}Q_{j',l} -i\Gamma_y \sum_{l'} \frac{e^{i \omega/c |y_l-y_l'|}}{\omega-\omega_0}Q_{j,l'}.
\end{equation}
By introducing the Green function $G=(\omega-H_{eff})^{-1}$, $Q_{j,l}$ can be written in a compact form of
\begin{equation}
    Q_{j,l}=\frac{g_x}{\sqrt{2\pi}}\sum_{j',l'} G_{j,l;j',l'} e^{i\kappa x_j'}f_{\kappa,l'}, \label{QijGreen}
\end{equation}
and we can consequently obtain
\begin{equation}
O_{k_x,l}=\delta_{k_x,\kappa}f_{\kappa,l}+\frac{g_x^2}{{2\pi}} \sum_{j,j',l'} \frac{e^{i(\kappa x_{j'}-k_x x_j)}}{\omega-\omega_{k_x}} G_{j,l;j',l'} f_{\kappa,l'}, \nonumber
\end{equation}
and 
\begin{equation}
W_{k_y,j}=\frac{g_xg_y}{{2\pi}} \sum_{l,j',l'} \frac{e^{-ik_y y_l}e^{i\kappa x_{j'}}}{\omega-\omega_{k_y}} G_{j,l;j',l'} f_{\kappa,l'}. \nonumber
\end{equation}
At last, the amptitudes of photon at the $y_l$ port in forward and backward scatterings can be calculated via
\begin{equation} \label{ForwardS}
    \chi_{x,l}=\int_{0}^{+\infty} O_{k_x,l} dk_x=f_{\kappa,l}-i\Gamma_x \sum_{j,j',l'} e^{-i\kappa {x_j}}e^{i\kappa x_{j'}} G_{j,l;j',l'} f_{\kappa,l'}, 
    \end{equation}
        and
\begin{equation}
\label{BackwardS}
    \chi_{\bar x,l}=\int_{-\infty}^{0} O_{k_x,l}dk_x
    =-i\Gamma_x \sum_{j,j',l'} e^{i\kappa x_j}e^{i\kappa x_{j'}} G_{j,l;j',l'} f_{\kappa,l'}, 
\end{equation}
while amplitude of photon at the $x_j$ port in upward and downward scatterings can be calculated via
\begin{equation}
      \chi_{y, j}=\int_{0}^{+\infty} W_{k_y,x_j} dk_y=-i\frac{g_xg_y}{c} \sum_{j',l,l'} {e^{-i\kappa y_l}e^{i\kappa x_{j'}}} G_{j,l;j',l'} f_{\kappa,l'},
         \end{equation}
         and
\begin{equation}
      \chi_{\bar y, j}=\int_{-\infty}^{0} W_{k_y,x_j} dk_y 
      =-i\frac{g_xg_y}{c} \sum_{j',l,l'} {e^{i\kappa y_l}e^{i\kappa x_j'}} G_{j,l;j',l'} f_{\kappa,l'}.
\end{equation}
The sum probability of all ports automatically satisfies $\sum_{v,l}|\chi_{v,l}|^2=1$, $v\in \left\{x, \bar x, y, \bar y\right\}$, indicating the conservation of photon number.  

We can also give the expression for photonic scattering under open boundary condition along the $x$ direction and periodic boundary condition along the $y$ direction.
The amplitudes of excitation should satisfy the periodic boundary condition along the $y$ direction, that is, $Q_{j,l}=Q_{j,l+N_y}$.
This relation leads to the quantization of quasi-momentum along the $y$ direction, $k_{y,n} =2\pi n/L_y$ with $n=1,2,..., N_y$ and $L_y=N_y d$.
We need to rewrite the integral of $k_y$ in Eq.~\eqref{Coupling} into the form of discrete summation, $\int d_{k_y}\rightarrow \frac{2\pi}{L_y} \sum_{n}$. 
Similarly, the motion equation for excitation becomes
\begin{equation}
    Q_{j,l}=\frac{g_x}{\sqrt{2\pi}}\frac{e^{i\kappa x_j}}{\omega-\omega_0}f_{\kappa,l}-i\Gamma_x \sum_{j'}\frac{e^{i \omega/c |x_j-x_j'|}}{\omega-\omega_0}Q_{j',l}
    +\frac{\Gamma_y}{L_y} \sum_{l',n} \frac{e^{i k_{y,n} (y_l-y_l')}}{(\omega-\omega_0)(|\kappa|-|k_{y,n}|)}Q_{j,l'}.
\end{equation}
Hence, we can derive the effective Hamiltonian under open boundary condition along the $x$ direction and periodic boundary condition along the $y$ direction,
\begin{equation} \label{EffHam}
    \tilde{H}_{eff}=\sum_{j,l}\omega_0 b_{j,l}^\dag b_{j,l}-i\Gamma_x  \sum_{j,l,j'} b_{j,l}^{\dag}b_{j',l}e^{i\omega/c|x_j-x_{j'}|}
    +\frac{\Gamma_y}{L_y}  \sum_{j,l,l'} b_{j,l}^{\dag}b_{j,l'} \sum_n\frac{e^{i\omega/c(y_l-y_{l'})}}{|\kappa|-|k_{y,n}|}.
\end{equation}
Here, by replacing the Green function as $G=(\omega-\tilde H_{eff})^{-1}$, we can  calculate forward and backward scattering along the $x$ direction via Eq.~\eqref{ForwardS} and Eq.~\eqref{BackwardS}, respectively.
To avoid possible divergence in numerical calculations,  a common wisdom is to add a tiny imaginary value in the denominator of $\tilde{H}_{eff}$.

\section{S3. Transfer matrix method }
Transfer matrix method is an important approach to solve problems concerning photon scattering in 1D WQED~\cite{ivchenko2013resonantNew}. 
Apparently, because of more emitted ports and complex connection of atomic arrays, it is more complicated to generalize the transfer matrix method to the 2D WQED.
In this section, we will present how to calculate the probability distribution of photon emitted from these ports in 2D WQED with the transfer matrix method.

The transfer matrix establishes relations among the scattering coefficients around individual atoms.
First, we need to change the effective Hamiltonian from the momentum space into the real space by making a Fourier transformation,
\begin{equation}
\begin{aligned}
    a_{B,l}(x)=&\frac{1}{\sqrt{2\pi}}\int_{-\infty}^0a_{k_x,l}e^{ik_xx}dk_x,\\
    a_{F,l}(x)=&\frac{1}{\sqrt{2\pi}}\int_{0}^{+\infty}a_{k_x,l}e^{ik_xx}dk_x,\\
    a_{D,j}(y)=&\frac{1}{\sqrt{2\pi}}\int_{-\infty}^0a_{k_y,j}e^{ik_yy}dk_y,\\
    a_{U,j}(y)=&\frac{1}{\sqrt{2\pi}}\int_{0}^{+\infty}a_{k_y,j}e^{ik_yy}dk_y,\\
\end{aligned}
\end{equation}
where $a_{B,l}(x)$ ($a_{F,l}(x)$)  annihilates a backward (forward) propagating photon along the $l$th horizontal waveguide at the position $x$, and $a_{D,j}(y)$ ($a_{U,j}(y)$)  annihilates a downward (upward) propagating photon along the $j$th vertical waveguide at the position $y$.
While the excitation part of the Hamiltonian does not change, we can obtain the interaction between photon and excitation and the photon part of Hamiltonian in real space as,
\begin{eqnarray}
    H_I^{R}=\sum_{j,l,\eta=\{F,B\}}\hbar g_x\int dx \delta_{(x-x_j)}a_{\eta,l}(x)b_{j,l}^\dag +\sum_{j,l,\eta=\{U,D\}}\hbar g_y \int dk_y \delta_{(y-y_l)}a_{\eta,j}(y)b_{j,l}^\dag+h.c., \nonumber
\end{eqnarray}
and
\begin{eqnarray}
  H_P^{R}&=&\sum_{l} i\hbar c\int dx[a_{B,l}^\dag(x){\partial_x}a_{B,l}(x)- a_{F,l}^\dag(x){\partial_x}a_{F,l}(x)] \nonumber \\
  &+&\sum_{j} i\hbar c\int dy[a_{D,j}^\dag(y)\partial_y a_{D,j}(y)- a_{U,j}^\dag(y){\partial_y}a_{U,j}(y)].\nonumber 
\end{eqnarray}

After obtaining the real-space Hamiltonian $H_R=H_A+H_P^{R}+H_I^{R}$, we need to solve the eigenvalue problem $H_R\ket{E_\omega} = E_\omega\ket{E_\omega}$ with $E_\omega =\hbar \omega$ and
\begin{equation}
\ket{E_\omega}=\sum_{j,l}e_{j,l}b_{j,l}^\dag\ket{0}+\sum_{l,\eta=\{F,B\}}\int dx \psi_{\eta,l}(x)a_{\eta,l}^\dag(x)\ket{0}
+\sum_{j,\eta=\{U,D\}}\int dy \Psi_{\eta,j}(y)a_{\eta,j}^\dag(y)\ket{0}.
\end{equation}
Here, $e_{j,l}$ is the probability amplitude of the excitation in the position ($x_j,y_l$). 
$\psi_{F,l}(x)$ ($\psi_{B,l}(x)$) are the probability amplitudes
for the forward (backward)-propagating photon at the position $x$ along the $l$th horizontal waveguide.
$\Psi_{U,j}(y)$ ($\Psi_{D,j}(y)$) are the probability amplitudes
for the upward (downward)-propagating photon at the position $y$ along the $j$th vertical waveguide.
By substituting the above ansatz into the Schr\"odinger equation with the real-space Hamiltonian $H_R$, we can obtain
\begin{align}\label{17}
    &(\omega_0-\omega)e_{j,l}+ g_x\sum_{\eta=\{F,B\}}\psi_{\eta,l}(x)+ g_y\sum_{\eta=\{U,D\}}\Psi_{\eta,j}(y)=0,\nonumber\\
    &(-i c\frac{\partial}{\partial x}-\omega)\psi_{F,l}(x)+\sum_j g_x \delta_{(x-x_j)}e_{j,l}=0,\nonumber \\
     &(i c\frac{\partial}{\partial x}-\omega)\psi_{B,l}(x)+\sum_j g_x \delta_{(x-x_j)}e_{j,l}=0,\\
      &(-i c\frac{\partial}{\partial y}-\omega)\Psi_{U,j}(y)+\sum_l g_y \delta_{(y-y_l)}e_{j,l}=0,\nonumber\\
      &(i c\frac{\partial}{\partial y}-\omega)\Psi_{D,j}(y)+\sum_l g_y \delta_{(y-y_l)}e_{j,l}=0.\nonumber
\end{align}

We can further assume the wavefunction of the propagating photon as plane waves with different amplitudes,

\begin{align}\label{16}
   \psi_{F,l}(x)=&\frac{e^{i\omega x/c}}{\sqrt{2\pi}}[\theta_{(x_1-x)}+t_{1,y_l}^2\theta_{(x_2-x)}\theta_{(x-x_1)}+t_{2,y_l}^3\theta_{(x_3-x)}\theta_{(x-x_2)}+\cdots+t_{N,y_l}^{N+1}\theta_{(x-x_N)}], \nonumber\\
   \psi_{B,l}(x)=&\frac{e^{-i\omega x/c}}{\sqrt{2\pi}}[r_{0,y_l}^{1}\theta_{(x_1-x)}+r_{1,y_l}^2\theta_{(x_2-x)}\theta_{(x-x_1)}+r_{2,y_l}^3\theta_{(x_3-x)}\theta_{(x-x_2)}+\cdots+t_{N-1,y_j}^{N}\theta_{(x-x_N)}], \nonumber\\
   \Psi_{U,j}(y)=&\frac{e^{i\omega y/c}}{\sqrt{2\pi}}[\theta_{(y_1-y)}+t_{1,x_j}^2\theta_{(y_2-y)}\theta_{(y-y_1)}+t_{2,x_j}^3\theta_{(y_3-y)}\theta_{(y-y_2)}+\cdots+t_{N,x_j}^{N+1}\theta_{(y-y_N)}],\\
   \Psi_{D,j}(y)=&\frac{e^{-i\omega y/c}}{\sqrt{2\pi}}[r_{0,x_j}^{1}\theta_{(y_1-y)}+r_{1,x_j}^2\theta_{(y_2-y)}\theta_{(y-y_1)}+r_{2,x_j}^3\theta_{(y_3-y)}\theta_{(y-y_2)}+\cdots+t_{N-1,x_j}^{N}\theta_{(y-y_N)}]. \nonumber
\end{align}

Here, $t_{m,x_j}^{m+1}(r_{m,x_j}^{m+1})$ is the transmission (reflection) coefficient between the $m$th and $(m + 1)$th atoms in the $j$th vertical waveguide, $t_{m,y_l}^{m+1}(r_{m,y_l}^{m+1})$ is the transmission (reflection) coefficient between the $m$th and $(m + 1)$th atoms in the $l$th horizontal waveguide.  
$\theta_{(x)}$ is the step function, 
\begin{equation}
\theta_{(x)}=\begin{cases}
1 & \text{if} ~x>0; \\
1/2 & \text{if} ~x=0; \\
0 & \text{if} ~x<0.
\end{cases}
\end{equation}
Focusing on the atom in the position ($x_j,y_l$), we substitute Eq.~\eqref{16} into Eq.~\eqref{17} and can further obtain
\begin{figure}
    \centering
    \includegraphics[width=0.5\linewidth]{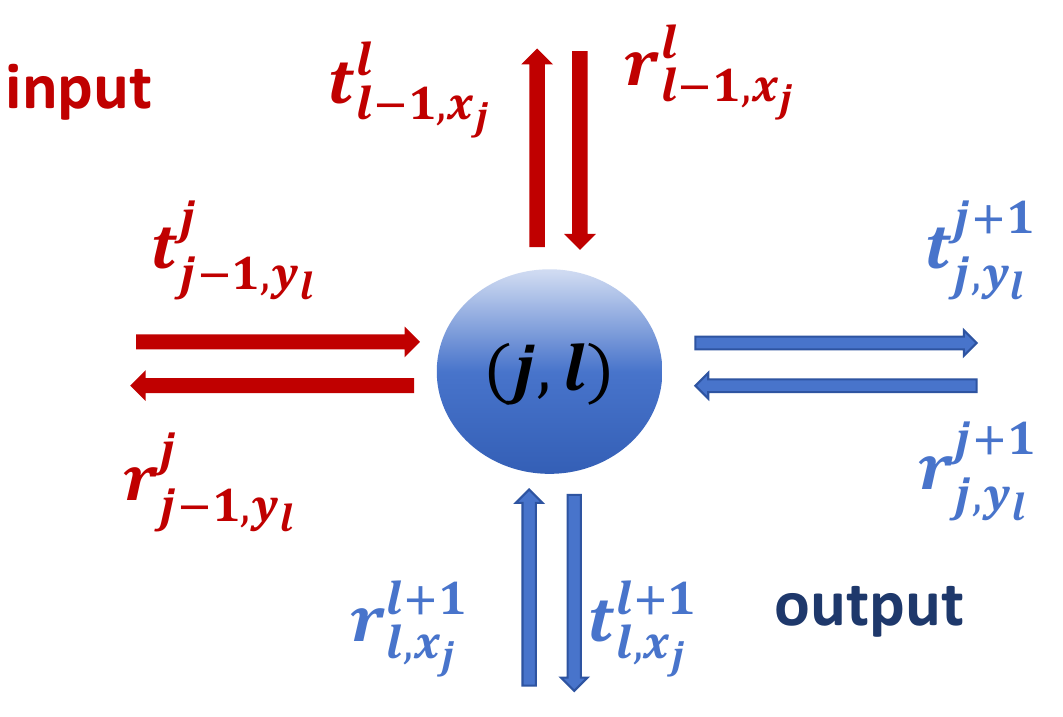}
    \caption{Schematics of transfer matrix of atom with position $(j,l)$, red part denote input parameters, blue part denote output parameters.}
    \label{transfer}
\end{figure}

\begin{equation}\label{18}
    \begin{aligned}
        -i\frac{e^{i\omega x_j/c}}{\sqrt{2\pi}}(-t_{j-1,y_l}^{j}+t_{j,y_l}^{j+1})+&\frac{g_x}{c}e_{j,l}=0,\\
        i\frac{e^{-i\omega x_j/c}}{\sqrt{2\pi}}(-r_{j-1,y_l}^{j}+r_{j,y_l}^{j+1})+&\frac{g_x}{c}e_{j,l}=0,\\
        -i\frac{e^{i\omega y_l/c}}{\sqrt{2\pi}}(-t_{l-1,x_j}^{l}+t_{l,x_j}^{l+1})+&\frac{g_y}{c}e_{j,l}=0,\\
        i\frac{e^{-i\omega y_l/c}}{\sqrt{2\pi}}(-r_{lj-1,x_j}^{l}+r_{l,x_j}^{l+1})+&\frac{g_y}{c}e_{j,l}=0,   
    \end{aligned}
\end{equation}
\begin{align}\label{19}
   \frac{\omega -\omega_0}{g_x}e_{j,l}&=\frac{e^{i\omega x_j/c}}{\sqrt{2\pi}}\frac{t_{j-1,y_l}^{j}+t_{j,y_l}^{j+1}}{2}+\frac{e^{-i\omega x_j/c}}{\sqrt{2\pi}}\frac{r_{j-1,y_l}^{j}+r_{j,y_l}^{j+1}}{2}
        \nonumber\\&+\lambda\frac{e^{i\omega y_l/c}}{\sqrt{2\pi}}\frac{t_{l-1,x_j}^{l}+t_{l,x_j}^{l+1}}{2}+\lambda \frac{e^{-i\omega y_l/c}}{\sqrt{2\pi}}\frac{r_{l-1,x_j}^{l}+r_{l,x_j}^{l+1}}{2},
\end{align}

where we set $\lambda =g_y/ g_x$ for brevity. By eliminating $e_{j,l}$ in Eqs.~\eqref{18} and \eqref{19}, we can build a relation between the reflection and transmission coefficients,
\begin{equation}\label{20}
    \begin{pmatrix}
        t_{j,y_l}^{j+1}\\
        r_{j,y_l}^{j+1}\\
        t_{l,x_j}^{l+1}\\
        t_{l,x_j}^{l+1}
    \end{pmatrix}=\mathcal M_{j,l}\begin{pmatrix}
        t_{j-1,y_l}^{j}\\
        r_{j-1,y_l}^{j}\\
        t_{l-1,x_j}^{l}\\
        t_{l-1,x_j}^{l}
    \end{pmatrix},
\end{equation}
where $\mathcal M_{j,l}=\mathcal A_{j,l}^{-1}\mathcal B_{j,l}$, $\mathcal A_{j,l}=\mathcal D+f_x\mathcal C_{j,l}$, $\mathcal B_{j,l}=\mathcal D-f_x \mathcal C_{j,l}$ with
\begin{equation}\label{21}
    \mathcal C_{j,l}=\begin{pmatrix}
    1&e^{-2i\kappa x_j}&\lambda e^{i\kappa(y_l-x_j)}&\lambda e^{-i\kappa(x_j+y_l)}\\
    e^{2i\kappa x_j}&1&\lambda e^{i\kappa(x_j+y_l)}&\lambda e^{i\kappa(x_j-y_l)}\\
    \lambda e^{i\kappa(x_j-y_l)}&\lambda e^{-i\kappa(x_j+y_l)}&\lambda^2&\lambda^2e^{-2i\kappa y_l}\\
    \lambda e^{i\kappa(x_j+y_l)}&\lambda e^{i\kappa(-x_j+y_l)}&\lambda^2 e^{2i\kappa y_l}&\lambda^2
    \end{pmatrix}, \nonumber
\end{equation}
$f_x=i\Gamma_x/(2(\omega_0-\omega)$, and $\mathcal D=\text{diag}(-1,1,-1,1)$. Fig.~\ref{transfer} shows the schematic diagram of the transfer matrix, in which the input coefficients in the left top of the $(j,l)$th atom are transferred to the output coefficients in the right bottom.  

Let us analyze the transfer matrix before we can actually apply it for scattering coefficients. 
The photon is initially injected from $y_0$ port and finally emitted from all possible ports.
The inputs from other $2(N_x+N_y)-1$ ports are $0$. 
From the initial condition, we have already known the $2(N_x+N_y)$ scattering coefficients.
There are total $4N_xN_y+2(N_x+N_y)$ scattering coefficients, which means that we need to determine $4N_xN_y$ unknown scattering coefficients. 
Because the transfer matrix at each atom gives $4$ linear equations, and the total $N_xN_y$ atoms can give $4N_xN_y$ linear equations. 
By solving these equations, we can uniquely and exactly determine all the scattering coefficients.

We can take a rectangular WQED composed of two horizontial and two vertical waveguides as an example. 
According to the initial condition, relations between all scattering coefficients are given by
\begin{equation}\label{24}
\begin{pmatrix}
    t_{1,y_1}^2\\
    r_{1,y_1}^2\\
    t_{1,x_1}^2\\
    r_{1,x_1}^2
\end{pmatrix}=\mathcal M_{1,1}\begin{pmatrix}
    1\\
    r_{0,y_1}^1\\
    0\\
    r_{0,x_1}^1
\end{pmatrix},
\end{equation}
\begin{equation}
\begin{pmatrix}
    t_{2,y_1}^3\\
    0\\
    t_{1,x_2}^2\\
    r_{1,x_2}^2
\end{pmatrix}=\mathcal M_{1,2}\begin{pmatrix}
    t_{1,y_1}^2\\
    r_{1,y_1}^2\\
    0\\
    r_{0,x_2}^1
\end{pmatrix},
\end{equation}
\begin{equation}
\begin{pmatrix}
    t_{1,y_2}^2\\
    r_{1,y_2}^2\\
    t_{2,x_1}^3\\
   0
\end{pmatrix}=\mathcal M_{2,1}\begin{pmatrix}
    0\\
    r_{0,y_2}^1\\
    t_{1,x_1}^2\\
    r_{1,x_1}^2
\end{pmatrix},
\end{equation}
\begin{equation}\label{27}
\begin{pmatrix}
    t_{2,y_2}^3\\
    0\\
    t_{2,x_2}^3\\
    0
\end{pmatrix}=\mathcal M_{2,2}\begin{pmatrix}
    t_{1,y_2}^2\\
    r_{1,y_2}^2\\
    t_{1,x_2}^2\\
    r_{1,x_2}^2
\end{pmatrix}.
\end{equation}
Combining the above $16$ linear equations, we can obtain exact values of the $16$ scattering coefficients.
From the above equations, we cannot obtain the final expression of scattering coefficients at the output ports in a simple form by iterating the transfer matrix just like the 1D WQED. 
We need to combine the transfer matrix equation for all the atoms in order to get the exact scattering probability distribution of individual ports.
The computational complexity will increase dramatically as the system size increases.
Besides, here we only apply the transfer matrix method to solve scattering problems in the rectangular lattice.
When the structure of an atomic array becomes more complex, the transfer matrix method becomes much more tedious and cumbersome.
%
\begin{figure*}
    \centering
    \includegraphics[width=0.98\textwidth]{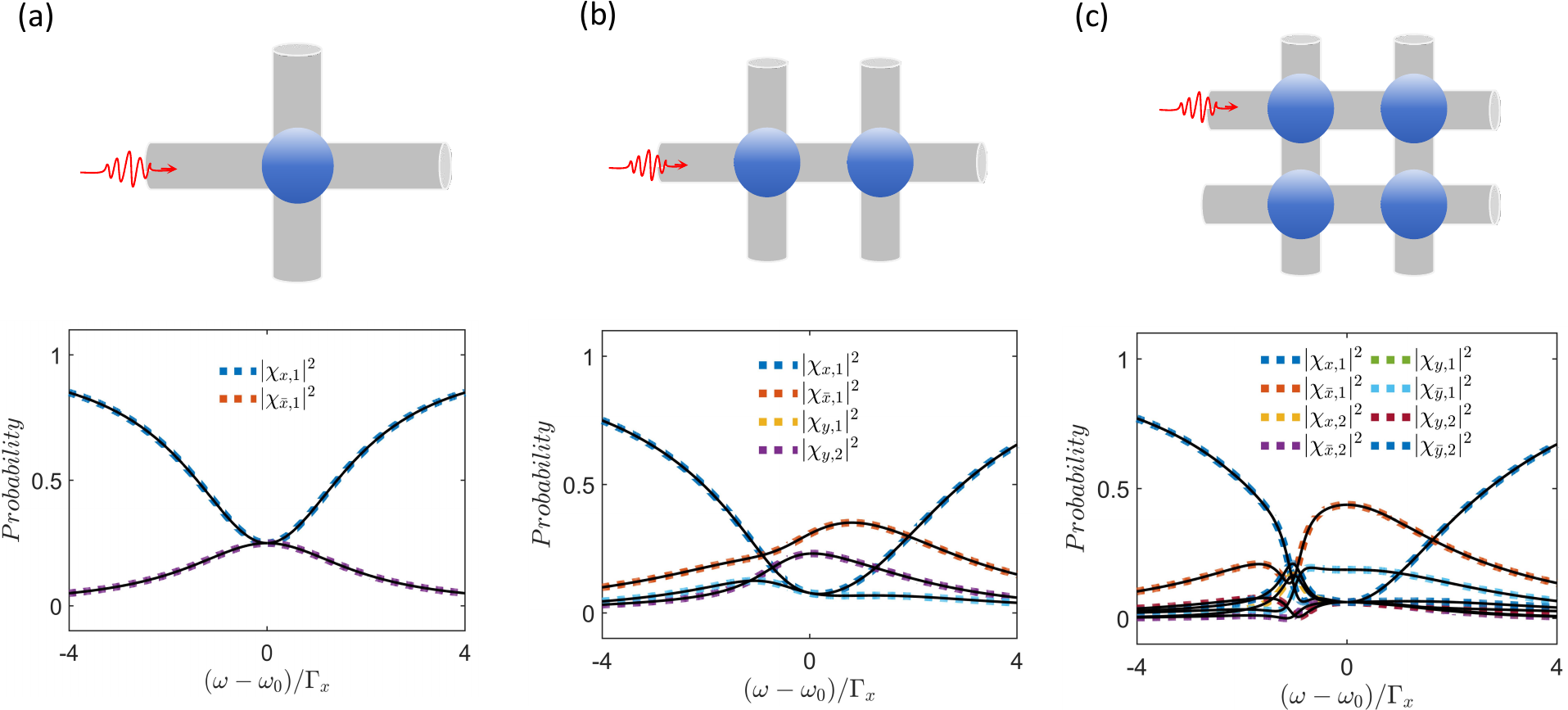}
    \caption{Scattering probability as a function of photonic frequency. (a) One atom. The curve of $|\chi_{y,1}|^2$, $|\chi_{\bar y ,1}|^2$ coincide with the curve of $|\chi_{\bar x,1}|^2$. (b) Two atoms. Yellow line coincides with the purple line, and green line coincides with the light blue line. (c) Four atoms at the nodes of two horizontal and two vertical waveguides. In the above cases, photon is injected from the first horizontal waveguide. Dash lines are obtained by Green function method, black solid lines are obtained transfer matrix method. The parameters are chosen $\varphi=\pi/6$, $c=100$, $\lambda=1$, $g_x=1$.
 }\label{f2}
\end{figure*}

\section{S4. Comparison between the Green function method and transfer matrix method}
We have presented the Green function method and the transfer matrix method for photonic scattering in 2D WQED.
In the following, we will show that the two methods can give exactly the same scattering probabilities of the output ports.
We consider three different structures: (i) one horizontal waveguide and one vertical waveguide, (ii) one horizontal waveguide and two vertical waveguide, (iii) two horizontal waveguide and two vertical waveguide.
Figs.~\ref{f2}(a-c) show the probability of photon emitted from some ports for the above three different structures, respectively.
The paramaters are chosen as $\varphi=\pi/6$, $c=100$, $\lambda=1$, and $g_x=1$.
The dashed lines represent the results obtained by the Green function method, and black solid lines represent the results obtained by the transfer matrix method. 
Some scattering probabilities are the same due to spatial symmetry of the system, such as
$\chi_{y,1}$ and $\chi_{\bar y,1}$, $\chi_{y,2}$ and $\chi_{\bar y,2}$ in Fig.~\ref{f2}(b).
The dashed lines and the black solid lines are perfectly consistent with each other, confirming the equivalence between the Green function method and the transfer matrix method.

Since the transfer matrix method is suitable for small and simple structures, this method can be treated as a good benchmark for developing new scattering theories in 2D WQED.
However, the transfer matrix method is not good at dealing with complex and large structures.
Compared with the transfer matrix method, the Green function method has a more compact and concise form and consumes less computation resource, which can be applied to arbitrarily complex structures at a large scale.

\section{S5. Decay and damped Oscillations} 

\begin{figure}[!htp]
    \centering
    \includegraphics[width=0.8\textwidth]{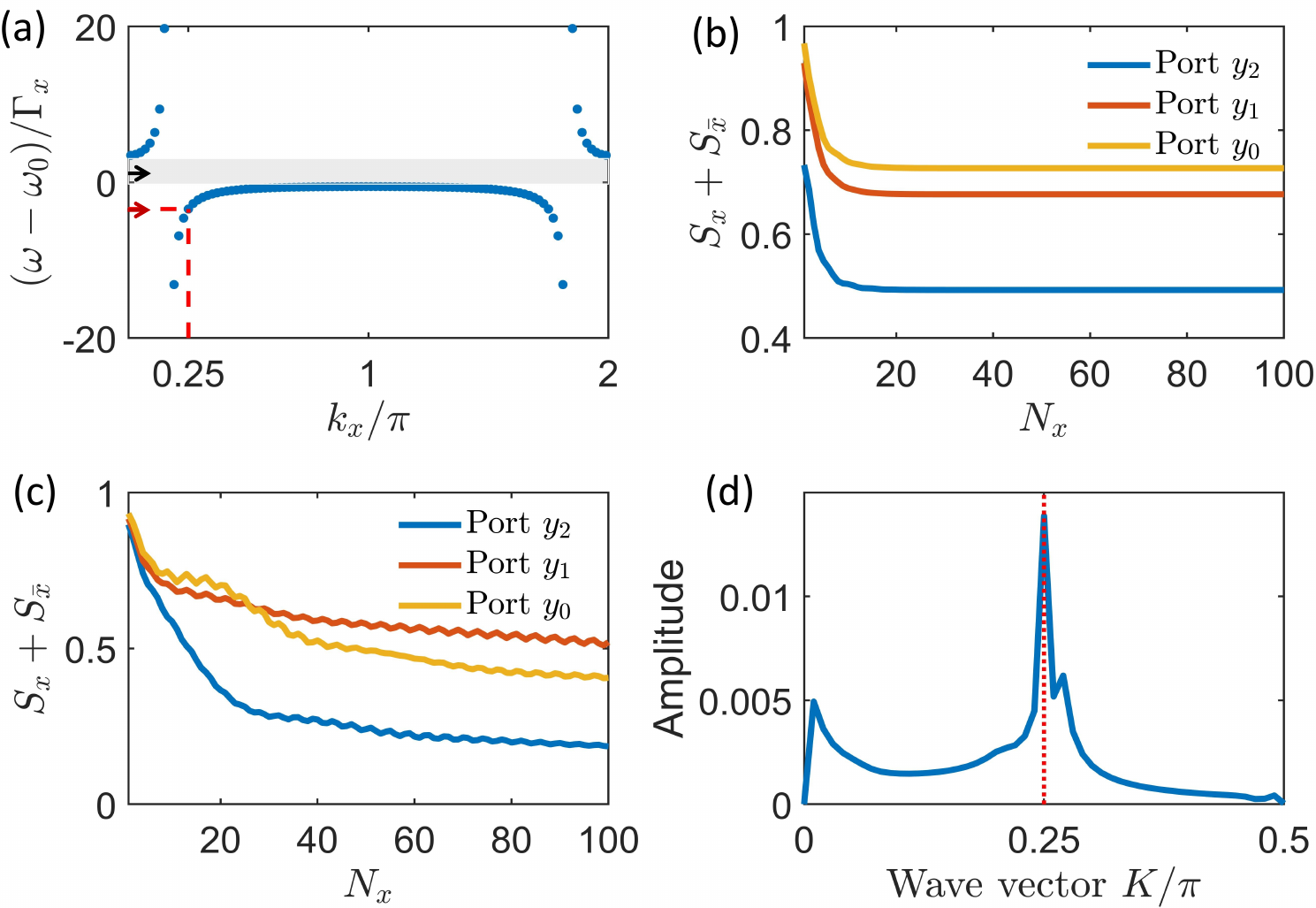}
    \caption{Oscillations and decay in horizontal scattering. (a) Energy band for ribbon structures with $N_y=5$ and $N_x=100$ under periodic (open) boundary condition along the $x$ ($y$) direction.  (b,c) Total probabilities of horizonal scattering as a function of $N_x$ with injected photonic frequencies $(\omega-\omega_0)/\Gamma_x=0.1821,-3.445$, respectively. The two frequencies are marked by black and red arrows in (a). (d) Fourier spectrum of (c). Blue, red, and green lines denote injected ports $y_2$, $y_1$ and $y_0$, respectively. Other parameters are chosen as $\varphi=\pi/6$, $g_x=g_y=1$, and $c=100$.
 }
    \label{SFig5}
\end{figure}
In this section, we try to study and explore the relationship between photonic scattering and the atomic layer.
In order to simplify the problem, we attempt to fix $N_y=5$ and observe how the horizontal scattering $S_x$ changes with $N_x$.
%
We first calculate the energy band; see Fig.~\ref{SFig5}.
There are $5$ nearly degenerate energy bands, which are separated as upper and lower branches with finite energy gaps.
We can set the frequency of injected photon in the energy gap or energy band.
For the photonic frequency in the gap [marked by the black arrow in Fig.~\ref{SFig5}(a)], we find that horizontal scattering quickly decays and then saturates to a certain value as $N_x$ increases; see Fig.~\ref{SFig5}(b).
Different injected ports have similar behavior but different saturated values.  
Because the photon always has chances to escape along the $y$ direction, and the probability of vertical scattering increases with the number of atomic layers along the $x$ direction.
However, the probability of vertical scattering does not increase to $1$, indicating that the ratio of vertical scattering to horizontal scattering converges to a certain value.
We will make this point clearer later.
However, for the photonic frequency in the energy band [marked by the red arrow in Fig.~\ref{SFig5}(a)], the horizontal scattering behaves as a damped oscillation with $N_x$ [Fig.~\ref{SFig5}(c)].
The oscillation is related to the resonant excitation of subradiant modes. 
We make a Fourier transformation of the horizontal scattering, and find that the amplitude as a function of wave vector ($K$) has the maximum peak at $K=\pi/4$; see Fig.~\ref{SFig5}(d).
The wave vector $K=\pi/4$ is exactly the same as the absolute momentum value of resonant subradiant states marked by the red arrow in Fig.~\ref{SFig5}(a).
%

\label{DecayOsc}

 \begin{table}
    \centering
    \resizebox{0.8\textwidth}{!}{
\begin{tabular}{ c|c|c|c|c|c|c|c|c|c|c  }
\hline
\hline
$(\omega -\omega_0)/\Gamma_x$&8.3442&14.0106&-9.0688&-3.445&-1.6642&-0.6642&-0.6160&-0.5967&-0.5821&-0.5775\\
\hline
$k_x$&$\pi/8$&$\pi/7$&$\pi/5$&$\pi/4$&$\pi/3$&$2\pi/3$&$3\pi/4$&$4\pi/5$&$6\pi/7$&$7\pi/8$\\
\hline
Period&7.7&7.143&5&4&3.03&3.03&4&5&7.143&8.33\\
\hline
\hline
\end{tabular}}
\caption{Spatial period of oscillation depends on frequency of rejected photon. $k_x$ is the corresponding wave vector of the injection frequency. The photon is injected from port $y_1$. Other parameters are chosen as $N_x=100$, and $N_y=5$.}
\label{t1}
\end{table}

%
For more general cases, When the frequency of an injected photon is in the energy band of excitation, the spatial period of the damped oscillations can be obtained by discrete Fourier transform; see table~\ref{t1}. 
Surprisingly, we find that the relation between spatial period and the quasi-momentum of the excitation approximately satisfies
\begin{equation}\label{48}
    D_x \approx \text{max}\left\{\frac{\pi}{k_x},\frac{\pi}{(\pi-k_x)}\right\},
\end{equation}
where $k_x$ is the $x$-direction quasi-momentum  of the most subradiant state in resonance with the injected photon. 
We can simplify 2D waveguide QED into 1D waveguide QED if we just consider the most subradiant state along the $y$ direction.
All scattering coefficients can be worked out with the transfer matrix method in 1D waveguide QED, which will help us to understand the decay and oscillation phenomena.

Because the decay rate of the most subradiant state along the $y$ direction is much smaller, at first we neglect the overall decay rate along the $y$ direction. 
According to Ref.~\cite{PhysRevLett.131.103604New}, we can obtain the expression of  transfer matrix corresponding to the atom with the position $x_j$,
\begin{equation}\label{49}
    \begin{aligned}
        M_j=\begin{pmatrix}
            -(1+2f_w)&-2f_we^{-2iz_j\omega/c}\\
           2f_we^{2iz_j\omega/c}&2f_\omega -1\\ 
        \end{pmatrix},
    \end{aligned}
\end{equation}
where $f_w=i\Gamma_x/(2\omega_0-2w)$, $z_j$ is the position of $j$th atom. We can establish the relation between the scattering coefficients $(t_{j-1}^{j},r_{j-1}^j)$ and $(t_{j}^{j+1},r_{j}^{j+1}) $ by transfer matrix $M_j$.
\begin{equation}\label{50}
    \begin{pmatrix}
        t_{j}^{j+1}\\
        r_{j}^{j+1}
    \end{pmatrix}=M_j\begin{pmatrix}
        t_{j-1}^{j}\\r_{j-1}^j
    \end{pmatrix}.
\end{equation}
By absorbing the accumulated phases in propagation of photons between two adjacent atoms~\cite{Liu_2023New}, we can redefine scattering coefficients 
\begin{eqnarray}    &\Tilde{t}_{j}^{j+1}&=e^{iz_j\omega/c}t_{j}^{j+1}, \nonumber\\
&\Tilde{r}_{j}^{j+1}&=e^{-iz_j\omega/c}r_{j}^{j+1},
\end{eqnarray}
which can be written by matrix form as following,
\begin{equation}\label{52}
   \begin{pmatrix}
        \Tilde{t}_{j}^{j+1}\\
        \Tilde{r}_{j}^{j+1} 
    \end{pmatrix}=\begin{pmatrix}
        e^{iz_j\omega/c}&0\\
        0&e^{-iz_j\omega/c}
    \end{pmatrix}\begin{pmatrix}
        t_{j}^{j+1}\\
        r_{j}^{j+1}
    \end{pmatrix}.
\end{equation}
Combining Eq.~\eqref{50} and Eq.~\eqref{52}, we can build the relationship between $(\Tilde{t}_{j-1}^{j},\Tilde{r}_{j-1}^j)$ and $(\Tilde{t}_{j}^{j+1},\Tilde{r}_{j}^{j+1}) $
\begin{equation}\label{53}
\begin{aligned}
    \begin{pmatrix}
        \Tilde{t}_{j}^{j+1}\\
        \Tilde{r}_{j}^{j+1} 
    \end{pmatrix}=&U_{j}M_jU_{j-1}^{-1}\begin{pmatrix}
            \Tilde{t}_{j-1}^{j}\\
            \Tilde{r}_{j-1}^{j}
    \end{pmatrix}\\
    =&P\begin{pmatrix}
            \Tilde{t}_{j-1}^{j}\\
            \Tilde{r}_{j-1}^{j}
    \end{pmatrix}
\end{aligned},
\end{equation}
where $P$ and $U_j$  are given by,
\begin{equation}\label{54}
    P=\begin{pmatrix}
        -(1+2f_w)e^{i\varphi}&-2f_we^{i\varphi}\\
        2fwe^{-i\varphi}&(2f_\omega -1)e^{-i\varphi}
    \end{pmatrix},
\end{equation}
\begin{equation}\label{55}
        U_{j}=\begin{pmatrix}
        e^{iz_j\omega/c}&0\\
        0&e^{-iz_j\omega/c}
    \end{pmatrix}.
\end{equation}
By iteratively applying the Eq.~\eqref{53}, we can obtain the relation between backward reflection $\Tilde{r}_{0}^1$ and forward transmission $\Tilde{t}_{N}^{N+1}$,
\begin{equation}\label{56}
   \begin{pmatrix}
        \Tilde{t}_{N}^{N+1}\\
        0 
    \end{pmatrix}= P^N\begin{pmatrix}
        1\\
        \Tilde{r}_{0}^1 
    \end{pmatrix}.
\end{equation}
We can diagonalize $P$ as $P=VQV^{\dag}$, where the diagonal elements of $Q$ are given by 
\begin{equation}\label{57}
    2f{\rm sin}\varphi-{\rm cos}\varphi \pm \sqrt{({\rm cos}\varphi-2f{\rm sin}\varphi)^2-1},
\end{equation}
with $f=\Gamma_x/(2\omega_0-2\omega)$. 
We consider the frequency of photon $\omega$ is in resonance with the excitation carrying quasi-momentum $k_x$ and energy
\begin{eqnarray}
\omega=\frac{\Gamma_x{\rm sin}\varphi}{{\rm cos}k_x-{\rm cos}\varphi}+\omega_0.
\end{eqnarray}
After a series of simplifications, the diagonal matrix $Q$ can be expressed with a very simple form,
\begin{equation}\label{59}
    Q=\begin{pmatrix}
        e^{ik_x}&0\\
        0&e^{-ik_x}
    \end{pmatrix}.
\end{equation}
Substituting Eq.~\eqref{59} and $P=VQV^{\dag}$ into Eq.~\eqref{56}, we can obtain
\begin{equation}\label{60}
  \begin{pmatrix}
        \Tilde{t}_{N}^{N+1}\\
        0 
    \end{pmatrix}=V\begin{pmatrix}
        e^{iNk_x}&0\\
        0&e^{-iNk_x}
        \end{pmatrix} V^{\dag}\begin{pmatrix}
            1\\
        \Tilde{r}_{0}^1 
    \end{pmatrix}.
\end{equation}
We rewrite Eq.~\eqref{60} and get the expression of $\Tilde{t}_{N}^{N+1}$,
\begin{equation}\label{61}
   t_N^{N+1}=\alpha(w,N)e^{iNk_x}+\beta(w,N) e^{-iNk_x}
\end{equation}
where $\alpha $ and $\beta$ are coefficients as functions of $\omega$ and $N$.
We can obtain transmission by taking the absolute value of both sides.
\begin{equation}\label{62}
 \begin{aligned}
     |t_N^{N+1}|^2=&\alpha^2+\beta^2+2\alpha\beta {\rm cos}(2Nk_x)\\
     =&\alpha^2+\beta^2+2\alpha\beta {\rm cos}(2N\pi-2Nk_x).\\
 \end{aligned}
 \end{equation}
 From Eq.~\eqref{62}, we know that transmission probability oscillates with system size. 
 Actually, the system size $N$ must be integer so that we cannot extract the oscillation period less than $1$. 
 Thus, the oscillation period should be the maximum value of $\left\{2\pi/2k_x, 2\pi/(2\pi-2k_x)\right\}$.
While the oscillation originates from the interference of resonant excitation states with momentum $k_x$ and $(2\pi-k_x)$,  the overall decay accompanying the oscillation comes from the decay of the most subradiant state along the $y$ direction.
 
 If the frequency of the injected photon is in the energy gap, the diagonal elements of $Q$ are two real numbers.
Moreover, one element is smaller than $1$ and the other is larger than $1$. 
We assume the eigenvalues of $Q$ are $e^{-\gamma}<1$ and $e^{\gamma}>1$. 
Similarly,  transmission probability  can be expressed as
\begin{equation}
|t_N^{N+1}|^2=\alpha^2e^{-2\gamma N}+\beta^2e^{2N\gamma}+\alpha \beta.
\end{equation}
Interestingly, we find the first term is dominant, $|t_N^{N+1}|^2\approx \alpha^2e^{-2\gamma N}$, indicating that the transmission probability decay as system size increases.
If $\omega$ tends to resonance frequency $\omega_0$,  the decay rate $\gamma$ will become much larger. 
We can obtain the range of energy gap $[-\tan(\varphi/2), \cot(\varphi/2)]$ by taking the boundary value of $k_x$, which can help us to judge whether oscillation or decay occurs.
Obviously, the decay rate in the case of pure decay is much larger than the one in the case of damped oscillation. That is because the pure decay has decay channels in both directions, whereas the damped oscillation has dominated decay channel in the $y$ direction.

\section{S6. Role of inversion symmetry in photonic scattering.} 
In the main text, we have observed that the total backward scattering is symmetric and the quantum GH shift in backward scattering is anti-symmetric with respect to the center port. 
The symmetric and anti-symmetric behaviors are determined by the inversion symmetry of the square lattice.
Here, we will strictly prove these relations. 

Using the eigenvalues $\{ \omega _n
\}$ and the eigenstates
$\ket{\{\psi_n}\}$ of the effective Hamiltonian (3) in the main text, the Green function can be written as 
\begin{equation}\label{green}
    G_{j,l;j',l'}(\omega _{\kappa})=\sum_n\frac{\psi_n(j,l)\psi_n(j',l')}{\omega _{\kappa}-\omega _n},
\end{equation}
where the eigenstate $\ket{\psi_n}$ has been normalized through 
\begin{eqnarray}
    \psi_n(j,l)=\frac{\psi_n(j,l)}{\sqrt{\sum_{j,l}\psi^2_n(j,l)}},
\end{eqnarray}
with $\psi_n(j,l)$ being the amplitude of $n$th eigenstate at the crossing of $j$th vertical waveguide and $l$th horizontal waveguide at the position $(x_j,y_l)$.
Because the couplings along the $x$ direction do not depend on the $y$ direction, or vise versa, we can further separate the wave functions as a product of the wave functions in the $x$ and $y$ directions,   
\begin{equation} \label{decomposition}
    {\psi_{(n_x,n_y)}(j,l)}=\psi_{n_x}(j)\psi_{n_y}(l).
\end{equation}
Here, two quantum numbers $n\equiv (n_x,n_y)$ are needed to uniquely determine a quantum state and the eigenvalue satisfies $\omega_n=\omega_{n_x}+\omega_{n_y}$.
Substituting Eq.~\eqref{decomposition} into Eq.~\eqref{green}, we can obtain
\begin{equation} \label{greenD}
        G_{j,l;j',l'}(\omega _{\kappa})=\sum_{n_x,n_y}\frac{\psi_{n_x}(j)\psi_{n_x}(j')\psi_{n_y}(l)\psi_{n_y}(l')}{\omega _{\kappa}-\omega_{n_x}-\omega_{n_y}}.
\end{equation}
By further substituting Eq.~\eqref{greenD} into Eq.~\eqref{BackwardS}, we can obtain the coefficients of backward scattering,
\begin{equation}
     \chi_{\bar x,l}=-i\Gamma_x \sum_{n_x,n_y} \frac{d_{n_x}^2 \psi_{n_y}(l)\psi_{n_y}(y_{in})}{\omega _{\kappa}-\omega_{n_x}-\omega_{n_y}},
\end{equation}
where $d_{n_x}=\sum_{j}e^{i\kappa x_j}\psi_{n_x}(j)$ is defined as the dipole moment of the $n_x$th eigenstate of the 1D WQED along the $x$ direction.
The probability of the total backward scattering is given by
\begin{equation}
    S_{\bar x}=  \sum_{n_x,n_y,n_x',n_y'} \frac{\Gamma_x^2 C_{n_y,n_y'}(y_{in}) d_{n_x}^2 (d_{n_x'}^{*})^2 }{(\omega _{\kappa}-\omega_{n_x}-\omega_{n_y})(\omega _{\kappa}-\omega_{n_x'}^{*}-\omega_{n_y'}^{*})}, \nonumber 
\end{equation}
where
\begin{equation}
    C_{n_y,n_y'}(y_{in})=\sum_l \psi_{n_y}(l) \psi_{n_y}(y_{in}) \psi_{n_y'}(l)^{*} \psi_{n_y'}(y_{in})^{*}
\end{equation}
is defined as the spatial correlation function between the $n_y$th and $n_y'$th eigenstates of the 1D WQED along the $y$ direction.

For simplicity, we denote the center position of the waveguide arrays along the $y$ direction as $0$.
Because of the inversion symmetry, and the eigenstates $\psi_{n_y}(l)$ satisfy $\psi_{n_y}(l)=\pm \psi_{n_y}(-l)$, where the plus and minus signs depend on the even and odd parities of the eigenstates, respectively.
Thus, $\psi_{n_y}(l)\psi_{n_y}(y_{in})= \psi_{n_y}(-l)\psi_{n_y}(-y_{in})$.
As a consequence, the spatial correlation satisfies
\begin{eqnarray}
    C_{n_y,n_y'}(-y_{in})&=&\sum_l \psi_{n_y}(l) \psi_{n_y}(-y_{in}) \psi_{n_y'}(l)^{*} \psi_{n_y'}(-y_{in})^{*} \nonumber \\
    &=& \sum_{-l} \psi_{n_y}(-l) \psi_{n_y}(-y_{in}) \psi_{n_y'}(-l)^{*} \psi_{n_y'}(-y_{in})^{*} \nonumber \\
    &=& \sum_{-l} \psi_{n_y}(l) \psi_{n_y}(y_{in}) \psi_{n_y'}(l)^{*} \psi_{n_y'}(y_{in})^{*} \nonumber \\
    &=& C_{n_y,n_y'}(y_{in}).
\end{eqnarray}
Because the spatial correlation is symmetric with respect to the center waveguide, we can immediately arrive at the conclusion that the probability of the total backward scattering remains the same if changing the injection port from $y_{in}$ to $-y_{in}$,
\begin{equation}
    S_{\bar x}(y_{in})=S_{\bar x}(-y_{in}).
\end{equation}
The above equation indicates that the probability of the total backward scattering is symmetric with respect to the center waveguide.

Below we will show that the mean position in the backward scattering is anti-symmetric with respect to the center waveguide.
Similarly, we can write the mean position in the backward scattering as 
\begin{equation}
P_{\bar x}=\frac{\Gamma_x^2}{S_{\bar x}}\sum_{n_x,n_y\\n_x',n_y'} \frac{ D_{n_y,n_y'}(y_{in}) d_{n_x}^2 (d_{n_x'}^{*})^2 }{(\omega _{\kappa}-\omega_{n_x}-\omega_{n_y})(\omega _{\kappa}-\omega_{n_x'}^{*}-\omega_{n_y'}^{*})}, \nonumber 
\end{equation}
where we define dipole correlation function as
\begin{equation}
    D_{n_y,n_y'}(y_{in})=\sum_l l \psi_{n_y}(l) \psi_{n_y}(y_{in}) \psi_{n_y'}(l)^{*} \psi_{n_y'}(y_{in})^{*}. \nonumber 
\end{equation}
By calculating dipole correlation function from the injection port $-y_{in}$, we can find that 
\begin{eqnarray}
     D_{n_y,n_y'}(-y_{in})
     &=&\sum_{-l} (-l) \psi_{n_y}(-l) \psi_{n_y}(-y_{in}) \psi_{n_y'}(-l)^{*} \psi_{n_y'}(-y_{in})^{*} \nonumber \\
     &=&\sum_{-l} (-l) \psi_{n_y}(-l) \psi_{n_y}(y_{in}) \psi_{n_y'}(l)^{*} \psi_{n_y'}(y_{in})^{*} \nonumber \\
     &=&-\sum_{l} l \psi_{n_y}(-l) \psi_{n_y}(y_{in}) \psi_{n_y'}(l)^{*} \psi_{n_y'}(y_{in})^{*} \nonumber \\
     &=& -D_{n_y,n_y'}(y_{in}), \nonumber
\end{eqnarray}
indicating the dipole correlation function is an odd function of injection port $y_{in}$.
Combining with the even function of $C_{n_y,n_y'}(y_{in})$, we can find that the mean position in the backward scattering is also an odd function of $y_{in}$,
\begin{equation}
    P_{\bar x}(-y_{in})=-P_{\bar x}(y_{in}).
\end{equation}
At last, the QGH shift in backward scattering satisfies
\begin{equation}
    P_{\bar x}(y_{in})-y_{in}=-[P_{\bar x}(-y_{in})-(-y_{in})],
\end{equation}
which is anti-symmetric about the center port. 
We can also prove the relation of QGH shifts between opposite injection ports $y_{in}$ and $-y_{in}$ in the forward scattering in the similar way.


%

\end{document}